\documentclass[aps,prd,preprint,showpacs,amsmath,amssymb,floatfix]{revtex4}
\pdfoutput=1
\usepackage{epsfig}
\usepackage{enumerate}
\usepackage{subfigure}
\usepackage{rotating}
\usepackage[bookmarks]{hyperref}
\usepackage{color}

\newcommand{\SLASH}[2]{\makebox[#2ex][l]{$#1$}/}
\newcommand{\pslash}{\SLASH{p}{.2}}

\newlength{\www}

\newcommand{\be}{\begin{equation}}
\newcommand{\ee}{\end{equation}}
\newcommand{\ba}{\begin{eqnarray}}
\newcommand{\ea}{\end{eqnarray}}

\newcommand{\bq}{\begin{equation}}
\newcommand{\eq}{\end{equation}}
\newcommand{\bqa}{\begin{eqnarray}}
\newcommand{\eqa}{\end{eqnarray}}
\newcommand{\ben}{\begin{enumerate}}
\newcommand{\een}{\end{enumerate}}
\newcommand{\bc}{\begin{center}}
\newcommand{\ec}{\end{center}}
\newcommand{\bqb}{\begin{eqnarray*}}
\newcommand{\eqb}{\end{eqnarray*}}

\newcommand{\mathh}{\mathcal{H}}

\newcommand{\pgsl}{p_g\hskip-0.35cm\slash~}
\newcommand{\qsl}{q\hskip-0.21cm\slash}
\newcommand{\qpsl}{q'\hskip-0.29cm\slash}
\newcommand{\esl}{\epsilon_g\hskip-0.33cm\slash (\mu)}

\newcommand{\drbar}{$\overline{\rm DR}$~}

\begin{document}

\title{\vspace{1cm}
Semi-inclusive bottom-Higgs production at LHC: \\ The complete one-loop electroweak effect in the MSSM
}
\author{
M.~Beccaria$^{a,b}$, 
G.~O.~Dovier$^{c,d}$,
G.~Macorini$^{a,b}$,
E.~Mirabella$^e$,
L.~Panizzi$^{f}$,
F.M.~Renard$^g$
and C.~Verzegnassi$^{c, d}$ \\
\vspace{0.4cm}
}

\affiliation{\small 
$^a$ Dipartimento di Fisica, Universit\`a del Salento, Italy\\
\vspace{0.2cm}
$^b$ INFN, Sezione di Lecce, Italy\\
\vspace{0.2cm}
$^c$ Dipartimento di Fisica, Universit\`a di Trieste, Italy\\
\vspace{0.2cm}
$^d$ INFN, Sezione di Trieste, Italy\\
\vspace{0.2cm}
$^e$  Institut de Physique Th\'eorique, CEA-Saclay,  France\\ 
\vspace{0.2cm}
$^f$  Institut de Physique Nucl\'eaire,
Universit\'e Lyon 1 and CNRS/IN2P3, France\\
\vspace{0.2cm}
$^g$ Laboratoire de Physique Th\'{e}orique et Astroparticules,
Universit\'{e} Montpellier II, France
}

\begin{abstract}
We present the first complete calculation of the one-loop electroweak effect in the
process of semi-inclusive bottom-Higgs production at LHC in the MSSM. The size
of the electroweak contribution depends on the choice of the final produced
neutral Higgs boson, and can be  relevant, 
in some range of the input parameters. A comparison of the
one-loop results obtained in two different renormalization schemes is also 
performed, showing a very good NLO scheme independence. We further comment on
two possible, simpler, approximations of the full NLO result, and on their reliabilty.
\end{abstract}



\maketitle

\section{Introduction}
\label{sec:intro}

It is a well known fact that $\tan \beta$ enhanced Yukawa coupling
in the Minimal Supersymmetryc Standard Model (MSSM) could favour the Higgs
production in association with bottom
quarks, contrarily to the Standard Model (SM) case, where the Higgs production is dominated
by top-Higgs coupling.\\

Due to its relevance as a possible channel for the Higgs discovery,
in the last few years the associated bottom-Higgs production has been
extensively  
studied in the literature.
Depending on the choice of the flavour-scheme in the partonic
description of the initial state and on the identified final state, one
can consider a number of different partonic sub-processes for $\mathh^0 +
b_{jets}$ production: while the choice of the 4 versus 5 flavour scheme is mainly
theoretically motivated, resulting in a reordering of the perturbative
expansion~\cite{Campbell:2004pu}, the requirement of a minimum number of tagged $b$ in the final
state is physically relevant in the signal extraction. Assuming the 5-flavour
scheme (which ensures a better convergence of the perturbative series resumming
large logarithms in the bottom PDF), 
one can consider three different types of production processes, depending on
the required final states:
the exclusive one where both bottom jets are tagged ($b \bar{b} \mathh^0$ final
state), the semi-inclusive one where only
one bottom quark is tagged ($b \mathh^0$), and 
the inclusive one where no bottom quark jets are tagged.
While the inclusive process has a larger cross section~\cite{EX-NLOQCD}, the
semi-inclusive with a high $p_{b,T}$ bottom in the final state is
experimentally more appealing~\cite{SI-NLOQCD1}.\\

The relative weights of the partonic processes
($b\bar{b}\to\mathcal{H}^0$, $bg\to b \mathcal{H}^0$, $gg\to b\bar{b}\mathcal{H}^0$)
are analyzed in~\cite{EX-NLOQCD}, where also the $\alpha_s$ corrections (NLO) to the
leading sub-process $b\bar{b}\to\mathcal{H}^0$ are computed. The NNLO order in
QCD ($\alpha_s^2$) for the same sub-process is calculated in~\cite{EX-NNLOQCD}, while the electroweak (SM
and MSSM) and SUSY-QCD NLO corrections have been computed in~\cite{I-EW},
showing that the size of electroweak corrections can be comparable, for large 
$\tan\beta$, with that of the strong ones.\\  

The associated semi-inclusive production process ($b \mathcal{H}^0$ final state) is analyzed at the
NLO in QCD in~\cite{SI-NLOQCD1} and~\cite{SI-NLOQCD2}, while the effect of the
SUSY QCD is given in~\cite{SI-SQCD}. Very recently, Dawson and Jaiswal have also computed, 
for the Standard Model process $bg \to b\,h_{SM}$, the one-loop weak
corrections~\cite{dj10}.\\ 

Finally, the exclusive process, where two bottom jets are tagged in the final
state, is considered at the NLO in QCD in~\cite{Campbell:2004pu},~\cite{djrw06},~\cite{Dittmaier:2003ej}
and~\cite{Dawson:2003kb}. The leading Yukawa corrections for this partonic process are considered
in~\cite{EX-EW} and SUSY QCD effects have also been computed in~\cite{EX-SQCD}.\\ 

Our paper is strongly motivated by the possible relevance of the associated
bottom-Higgs production in the experimental search of the Higgs at the LHC;
moreover, as 
stressed in~\cite{I-EW}, the susy one-loop ew effects (for the inclusive 
process) can be sizable and they can be safely accounted by an improved 
born approximation. Therefore the spirit of our computation is twofold: 
on the one hand we provide for the first time the complete NLO EW 
corrections for the semi-inclusive process, including also the overall 
QED effect, that was not computed by~\cite{dj10}, and on the other hand we can perform a further and 
independent test on the validity and limits of the improved born 
approximation in different scenarios.
Our calculations have been 
performed in  two different (\drbar and DCPR) renormalization schemes:
as expected the final one-loop results are, within at most a relative few percent
difference, the same in the two frames; however, 
 the \drbar scheme appears to be the one where the perturbative effect is 
numerically mostly more under control.
Therefore we shall discuss  our  results in this frame, showing
in various figures the dependence of the different observables on the choice of the
input parameters. We have finally compared the results obtained with the full
electroweak computation with those obtained within a commonly used
approximation scheme. This will be done in the final part of our paper, which  is
organized as follows: Section~\ref{sec:kin} contains a general concentrated
discussion of the actual derivation of the theoretical formulae (a part of which
has been shifted in a technical Appendix~\ref{app:A}) to be used for the calculation of the
various observables. Section~\ref{sec:num} and~\ref{sec:IBA} contains our
numerical results, that are briefly discussed  in the final Section~\ref{sec:concl}.

\section{Kinematics and Amplitude of the process $bg\to b\mathh^0$}
\label{sec:kin}

\subsection{Kinematics}
\label{sub:A}
At lowest order there is only one partonic$^{\mbox{\tiny{\ref{foot:pb}}}}$ channel leading to bottom-Higgs production
\footnotetext[1]{One should also consider
  the photon induced process $b\gamma \to b\mathh^0$: the contribution to the
  total cross section arising from this sub-proceess is doubly suppressed, due
  to the smaller $\gamma$ parton distribution function  and smaller coupling ($\alpha$
  instead of $\alpha_s$). Since
  the main goal of this paper is the calculation of the NLO electroweak  effects for
  $b\mathh^0$ production, and
  the $b\gamma \to b\mathh^0$ can be safetly computed at the LO, we do not take
  into account the photon induced production in the following.
  \label{foot:pb}}
\bq
b(p_b)~g(p_g)\to b(p'_b)~\mathcal{H}^{0}(p_{\mathh^0})
\eq
where $\mathcal{H}^0$
is one of the three MSSM neutral Higgs bosons ($h^0,\, H^0,\, A^0$).
In the partonic center of mass frame
the momenta of the particles read
\begin{eqnarray}
  p_b = (E_b;0,0,p), &~~& p_g=(p;0,0,-p), \nonumber \\
  p'_{b}=(E'_{b};p'\sin\theta,0,p'\cos\theta), &~~& p_{\mathh^0}=(E_{\mathh^0};-p'\sin\theta,0,-p'\cos\theta).
\end{eqnarray}
The Mandelstam variables are defined as
\bq
\label{Eq:Mand}
s = (p_g+p_b)^2,~~ t =(p_{b}-p'_b)^2 ,~~ u=(p_g-p'_b)^2.
\eq
For later convenience we define two momenta $q$ and $q'$ as follows
\begin{displaymath}
q=p_b + p_g ,~~~~~q'=p'_b - p_g
\end{displaymath}

\subsection{Born and one-loop amplitudes}
\label{sub:B}

\begin{figure}
\centering
\includegraphics[width=0.5\textwidth]{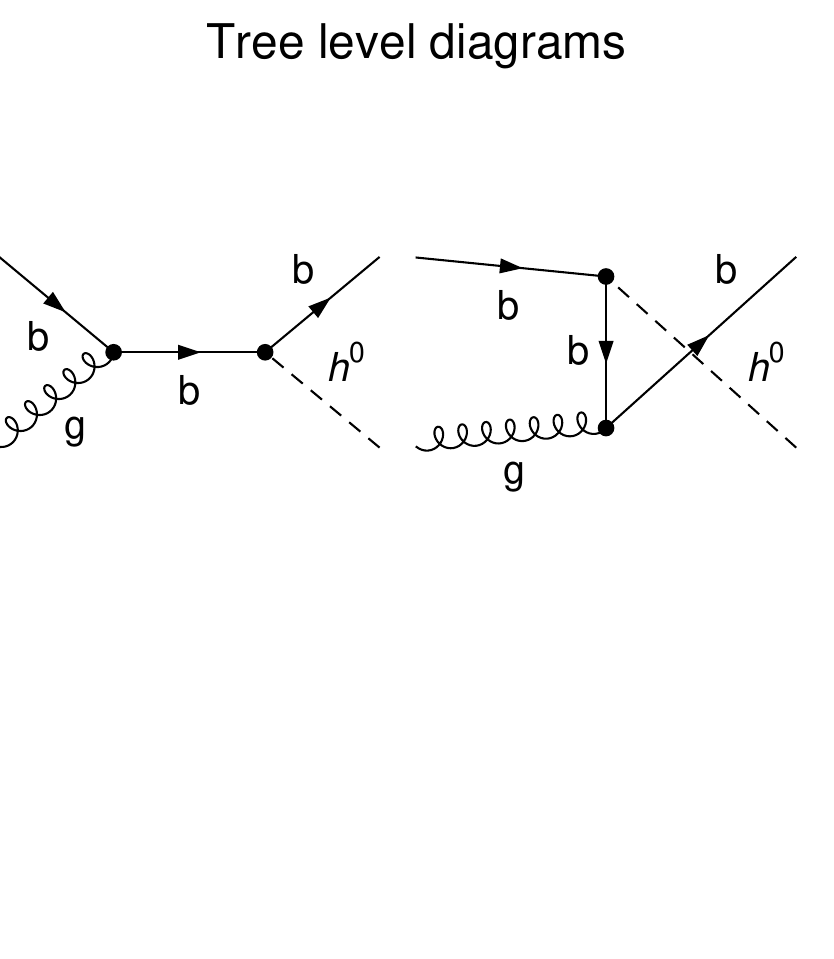}
\caption{Tree level diagrams for the partonic $bg \to b\mathh^0$ processes.
\label{fig:tree}}
\end{figure}

\begin{figure}
\centering
\includegraphics[width=0.4\textwidth]{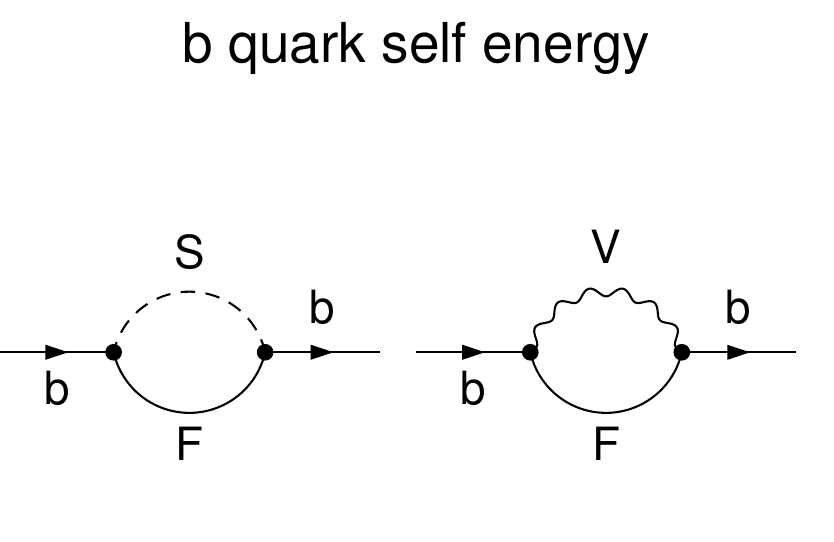}
\includegraphics[width=0.7\textwidth]{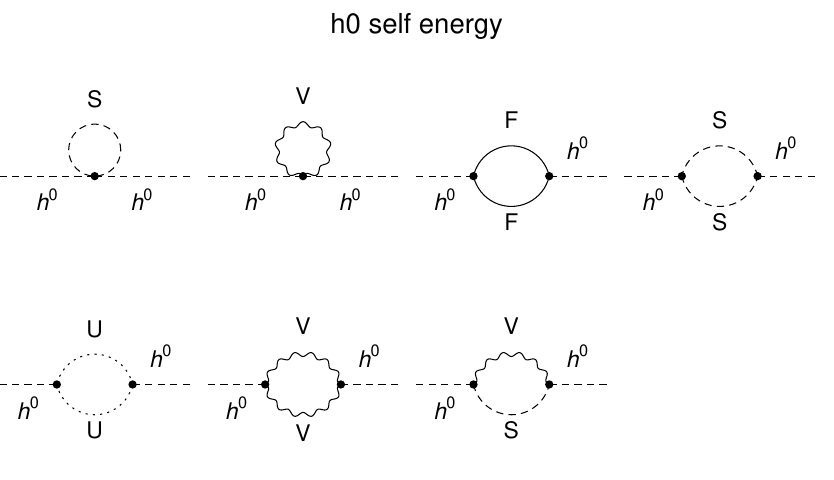}
\includegraphics[width=0.6\textwidth]{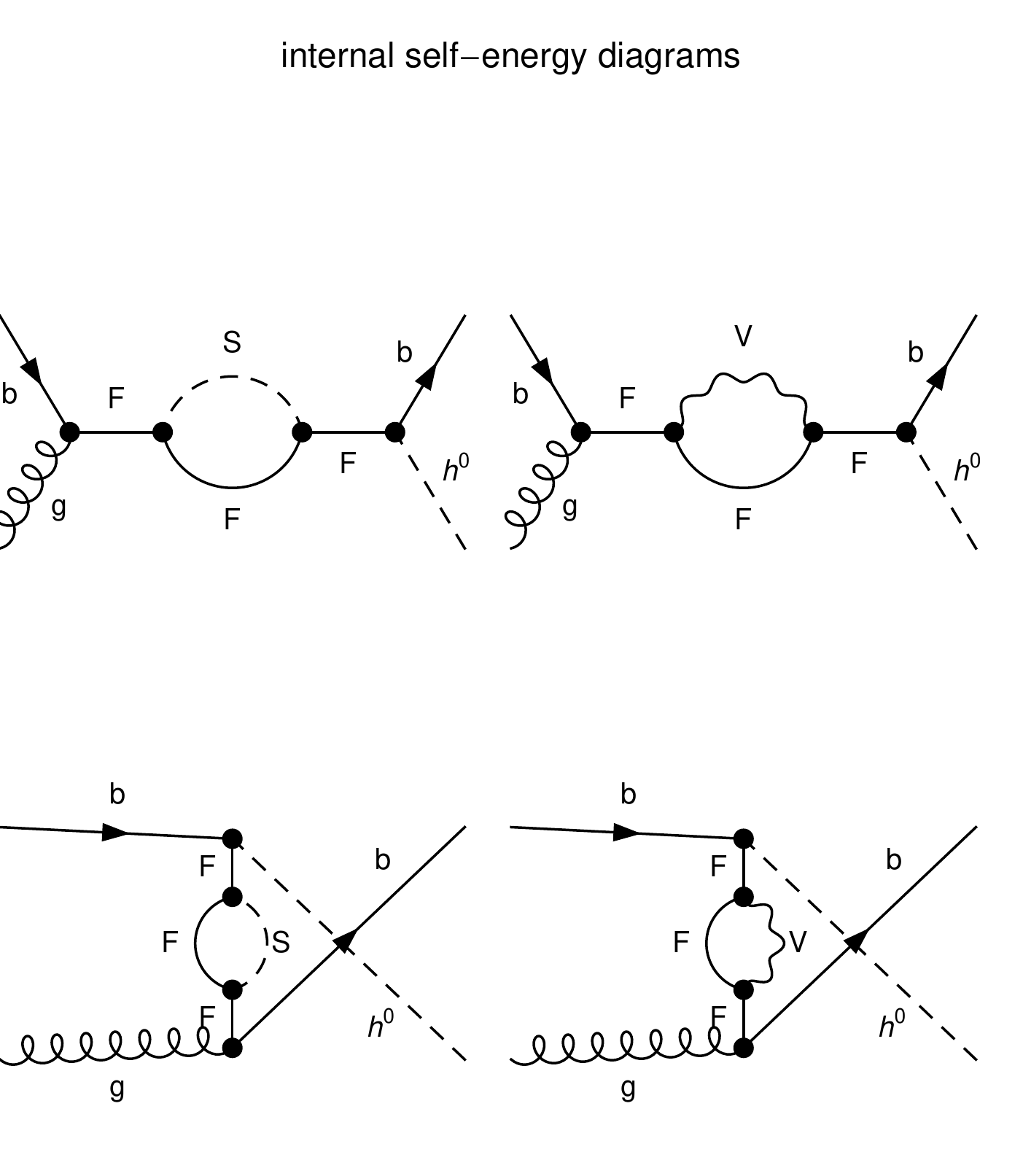}
\caption{Bottom quark self energies, higgs self energies (only the diagonal
  case) and internal self energies.
\label{fig:s.e.}}
\end{figure}

\begin{figure}
\centering
\includegraphics[width=0.45\textwidth]{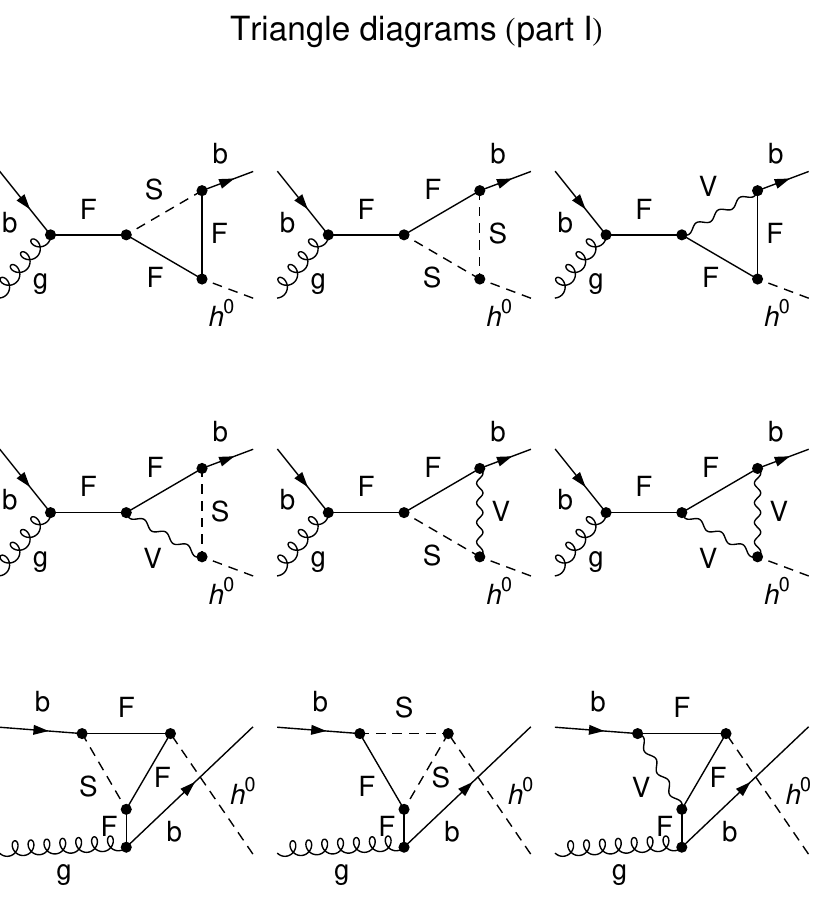}
\includegraphics[width=0.45\textwidth]{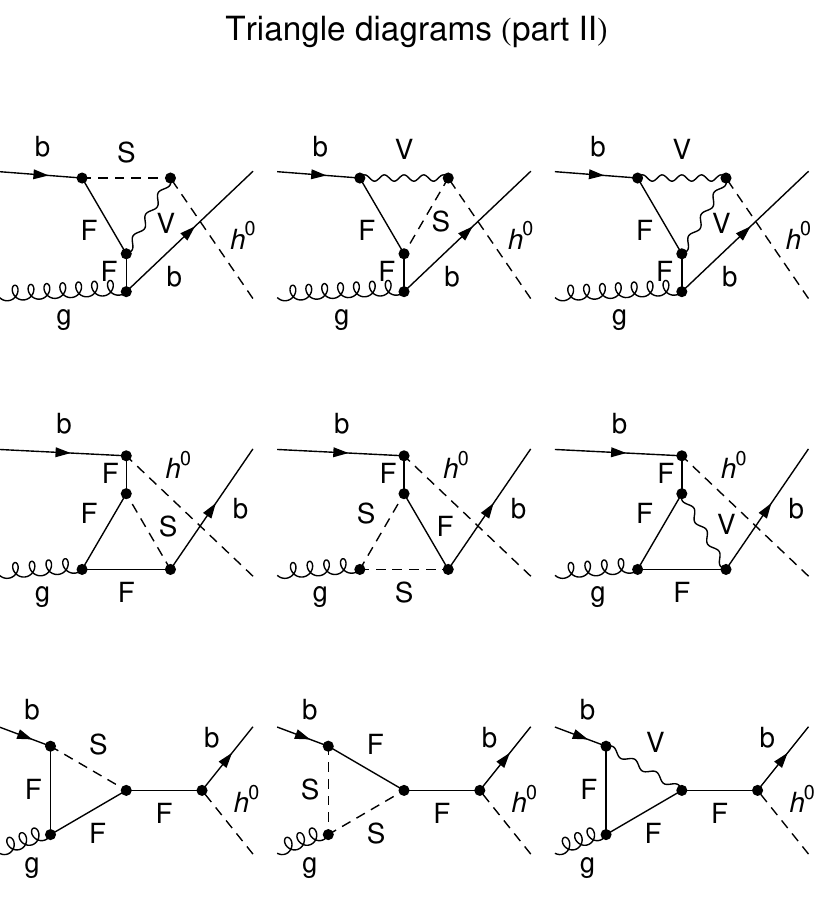}
\includegraphics[width=0.65\textwidth]{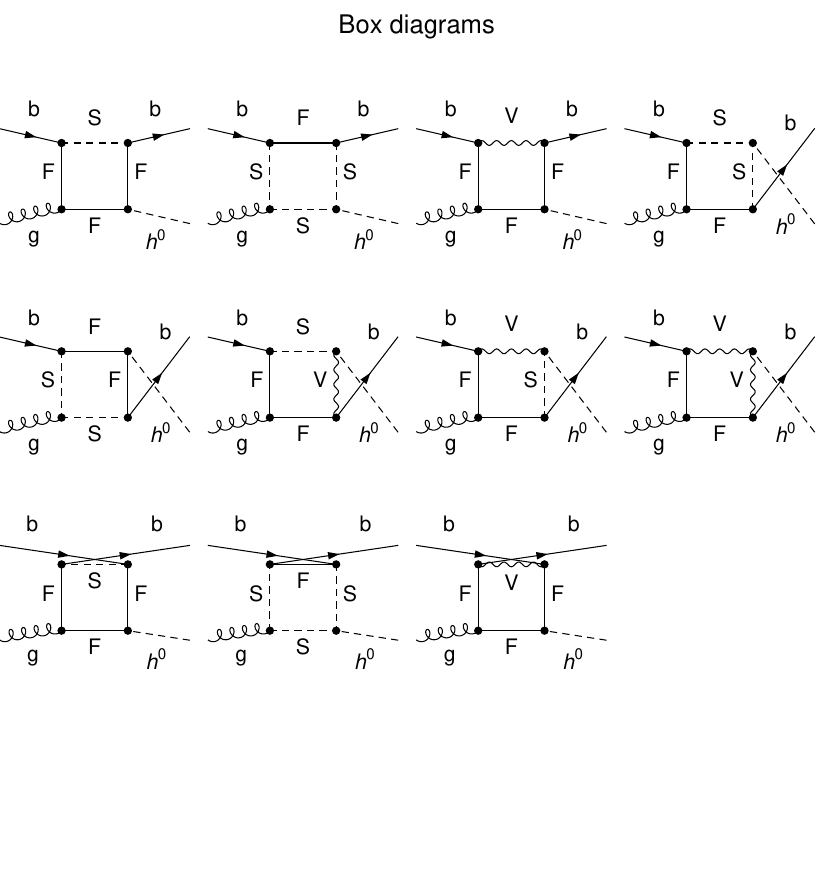}
\caption{Triangle and box diagrams.
\label{fig:tre_e_box}}
\end{figure}

We denote the $\mathcal{O}(\alpha^a_s \alpha^b)$ contribution to the
amplitude (differential cross section)  of the process $X$ as
$\mathcal{M}^{a,b}_X$ ($d \sigma^{a b}_X$).
The Born terms result from the $s$- and $u$-channel $b$ quark exchange of
Figure~\ref{fig:tree}.
The color stripped tree-level amplitude reads as follows
\begin{eqnarray} 
\mathcal{M}^{1/2,1/2}_{bg\to b \mathh^0}
&=&-\left ({g_s\over s-m^2_b} \right )
\bar u'_b(\lambda'_b)
[c^L(bb\mathh^0)P_L+c^{R}(bb\mathh^0)P_R]
(\qsl+m_b)\esl u_b(\lambda_b),\nonumber\\
&~& -  \left ({g_s\over u-m^2_{b}}\right )
\bar u'_b(\lambda_b')\esl(\qpsl+m_b)
[c^L(bb\mathh^0)P_L+c^{R}(bb\mathh^0)P_R]u_b(\lambda_b), 
\label{Eq:amplitude}
\end{eqnarray}
where $\lambda_b$, ($\lambda'_b$) is the helicity of the initial (final)
bottom quark while $\mu$ is the polarization of the gluon. 
$u_b(\lambda_b)$ [$u'_b(\lambda'_b)$] is the spinor of the initial [final]
bottom quark,  $\epsilon_g(\mu)=(0; \mu / \sqrt{2},-~i  / \sqrt{2},0)$  is 
the gluon polarization vector and $P_{R,L}=(1\pm\gamma^5)/2$ are the chirality
projectors. The relevant couplings 
$c^{\eta}(bb\mathh^0)$ ($\eta=L,R$) are defined as 
\begin{eqnarray}
  c^\eta(bbH^0)=-~\left ({em_b\over2s_WM_W} \right ){\cos\alpha\over\cos\beta},&~~& 
  c^\eta(bbh^0) =\left ({em_b\over2s_WM_W}\right ){\sin\alpha\over\cos\beta} \nonumber\\
  c^L(bbA^0)=
  - i \left ({em_b\over2s_WM_W} \right ) \tan\beta, &~~& c^R(bbA^0)= c^{L*}(bbA^0)~.
  \label{eq:couplings}
\end{eqnarray}
We factorize out of
the gluon couplings the colour matrix element  $\lambda^a  / 2$. The sum over colors leads to a  factor
\bq
\sum_{a=1}^8
\mbox{tr}\left \{  \frac{\lambda^a}{2}~\frac{\lambda^a}{2} \right \}  = 4  
\eq
that multiplies the squared amplitude.\\
The generic helicity amplitude 
can be decomposed on a set of eight forms factors 
$J^{k\eta}$ ($\eta = L,R$) as follows
\be
\mathcal{M}^{1/2,1/2}_{bg\to b \mathh^0}
= \bar u'_b(\lambda'_b) \left ( \sum_{k=1}^4 \sum_{\eta =L,R} 
J^{k\eta} N^{k\eta}_{bg\to b \mathh^0}
 \right ) u_b(\lambda_b),
\ee
where 
\bq
\label{eq:JFF}
J_{1\eta}=\pgsl\esl P_{\eta},~~~
J_{2\eta}=(\epsilon_g(\mu) .p'_{b}) P_{\eta},~~~
J_{3\eta}=\esl P_{\eta},~~~
J_{4\eta}=(\epsilon_g(\mu) .p'_{b})\pgsl P_{\eta}.
\eq
The only non-zero scalar functions at the tree level 
are $N^{1\eta}_{bg\to b \mathh^0}$ and 
$N^{2\eta}_{bg\to b \mathh^0}$. They read as follows
\bq
N^{1\eta}_{bg\to b \mathh^0} =-g_s {c^{\eta}(bb\mathh^0)\over s-m^2_b}-g_s
{c^{\eta}(bb\mathh^0)\over u-m^2_{b}},~~~
N^{2\eta}_{bg\to b \mathh^0} =-2g_s
{c^{\eta}(bb\mathh^0)\over u-m^2_{b}}
\eq
The one-loop electroweak virtual contributions arise from self energy, vertex and box
diagrams. Counterterms for the various bottom quarks lines, for the
$\mathcal{H}^0$ line, and for the $bb\mathh^0$ coupling constants have to be
considered as well. The corresponding diagrams can be read off from Fig~\ref{fig:s.e.},~\ref{fig:tre_e_box}.\\
All these contributions have been computed using the
usual decomposition in  terms of Passarino-Veltman functions and
the complete amplitude has been implemented in a C++ numerical code.\\

\subsection{Renormalization}
\label{sub:C}
In order to cancel the ultraviolet (UV) divergences the Higgs sector and the bottom sector
have to be renormalized at $\mathcal{O}(\alpha)$. The expressions of the  counterterms
entering our calculation are collected in Appendix~\ref{app:A}.\\

\noindent
{\bf Higgs sector}\\
As anticipated we performed the calculation using two different
renormalization schemes: the $\overline{\mbox{DR}}$ scheme \cite{Frank:2006yh} is defined by the
following renormalization conditions  
\begin{eqnarray}
  \delta Z^{\overline{\mbox{\tiny DR}}}_{H_1} &=& -\left [\,\mbox{Re}
    \frac{\partial \Sigma_{H^0}(k^2)}{\partial k^2}|_{k^2=M^2_{H^0}, \alpha=0}\,
  \right ]_{\mbox{\tiny div}}\, \nonumber \\
  \delta Z^{\overline{\mbox{\tiny DR}}}_{H_2} &=& -
  \left [\,\mbox{Re} \frac{\partial \Sigma_{h^0}(k^2)}{\partial
      k^2}|_{k^2=M_{h^0},\alpha=0}\,\right ]_{\mbox{\tiny div}}\, \nonumber \\
  \delta T_{h^0} &=&-T_{h^0}\, \nonumber \\
  \delta T_{H^0}&=&-T_{H^0}\, \nonumber \\
  \delta M_{A^0}^2 &=& \mbox{Re} \Sigma_{A^0}(M_{A^0}^2) - M_{A^0}^2  \Sigma'_{A^0}(M_{A^0}^2) \nonumber \\
  \delta\tan\beta^{\overline{\mbox{\tiny DR}}}&=&\frac{1}{2}\left (
    \delta Z^{\overline{\mbox{\tiny DR}}}_{H_2}-\delta
    Z^{\overline{\mbox{\tiny DR}}}_{H_1}\right ) \tan\beta.
\end{eqnarray}
$\delta Z^{\overline{\mbox{\tiny DR}}}_{H_i}$ define the wave
function renormalization costant of the Higgs field $H_i$,
the third and fourth line fix the tadpole
renormalization and the last  one the $\tan \beta$ renormalization constant.
$[\mathcal{A}]_{\mbox{\tiny div}}$ means keeping the UV divergent part of 
$\mathcal{A}$, discarding the finite contribution.
In the  DCPR scheme~\cite{Dabelstein:1994hb, Chankowski:1992er} the independent parameters are the same, and  
the renormalization conditions of the Higgs wavefunctions change as follows 
\begin{eqnarray}
  \delta Z^{\mbox{\tiny DCPR}}_{H_1} &=&
  -\mbox{Re} \frac{\partial \Sigma_{A^0}(k^2)}{\partial k^2}|_{k^2=M^2_{A^0}}
  - \frac{1}{\tan\beta M_Z} \mbox{Re} \Sigma_{A^0Z}(M^2_{A^0})\, \nonumber \\
  \delta Z^{\mbox{\tiny DCPR}}_{H_2} &=&
  -\mbox{Re} \frac{\partial \Sigma_{A^0}(k^2)}{\partial k^2}|_{k^2=M^2_{A^0}}
  + \frac{\tan \beta}{M_Z} \mbox{Re} \Sigma_{A^0Z}(M^2_{A^0})\, \nonumber \\
  \delta T_{h^0} &=&-T_{h^0}\,  \nonumber \\ 
  \delta T_{H^0} &=& -T_{H^0}\, \nonumber\\
  \delta M_{A^0}^2 &=& \mbox{Re} \Sigma_{A^0}(M_{A^0}^2) - M_{A^0}^2  \Sigma'_{A^0}(M_{A^0}^2) \nonumber \\
  \delta\tan\beta^{\mbox{\tiny DCPR}}&=&\frac{1}{2}\left (\delta Z^{\mbox{\tiny DCPR}}_{H_2}-\delta Z^{\mbox{\tiny DCPR}}_{H_1}\right )\tan\beta 
\end{eqnarray}
We choose to impose on-shell (OS) condition for the mass
of CP-odd $A^0$ Higgs in both schemes.\\

The renormalization constants of the Higgs bosons wavefunctions and 
of the $c^\eta(bb \mathh^0)$ couplings can be written in terms of the
of the renormalization constants defined above. Their explicit expression 
is   given in Appendix~\ref{App:AA}. \\

\noindent
{\bf Bottom sector}\\
The mass of the bottom and its wavefunction renormalization function is fixed in the on-shell
scheme:
\begin{eqnarray}
  \delta m^{\mbox{\tiny OS}}_b &=& \frac{1}{2} m_b \left [\mbox{Re}
    \Sigma_{{b}_L} (m_b^2) +
    \mbox{Re}\Sigma_{{b}_R} (m_b^2) + 
    2  \mbox{Re} \Sigma_{{b}_S} (m_b^2) \right ], \\
  \delta Z^L_b &=& -\mbox{Re} \Sigma_{{b}_L} (m_b^2) - m_b^2
  \frac{\partial}{\partial k^2} \mbox{Re} \left [ \Sigma_{{b}_L}(k^2)
    +\Sigma_{{b}_R}(k^2) +2\Sigma_{{b}_S} (k^2)   \right ]|_{k^2=m_b^2},
  \nonumber\\
  \delta Z^R_b &=& -\mbox{Re} \Sigma_{{b}_R} (m_b^2) - m_b^2
  \frac{\partial}{\partial k^2} \mbox{Re} \left [ \Sigma_{{b}_L}(k^2)
    +\Sigma_{{b}_R}(k^2) +2\Sigma_{{b}_S} (k^2)   \right ]|_{k^2=m_b^2},\nonumber
\end{eqnarray}
where the bottom self energies are defined according to following Lorentz
decomposition:
\begin{equation}
  \Sigma_b (p) = \pslash P_L
  \Sigma_{{b}_L} (p^2) + \pslash P_R
  \Sigma_{{b}_R} (p^2) + m_b \Sigma_{{b}_S} (p^2)\; .
\end{equation}
The bottom masses in the Yukawa couplings are treated completely at the
electroweak level, with OS or  $\overline{\mbox{DR}}$ renormalization conditions respectively in the two
schemes.  
Resummation of large logarithms from the running of the bottom mass  suggests to trade 
bottom mass appearing in the couplings with an effective 
bottom mass, 
~\cite{Heinemeyer:2004xw}. 
The resummation of the  $(\alpha_s \tan\beta)^n$ contributions can be achieved  
modifying the tree level relation between the bottom Yukawa coupling and
the bottom mass: the 
bottom mass of the couplings,
which is related to the bottom Yukawa coupling, 
is replaced by an effective mass, (e. g. in the $\overline{\mbox{DR}}$ scheme) 
\begin{equation}
\label{Eq:MBRES}
  m^{\overline{\mbox{\tiny DR}}}_b \to
  \overline{m}^{\overline{\mbox{\tiny DR}}}_b = \frac{m^{\overline{\mbox{\tiny
          DR}}}_b}{1+\Delta_b}
\end{equation}
where $\Delta_b$ is given by
\begin{eqnarray}
  \Delta_b &=& \frac{2}{3}\frac{\alpha_s}{\pi}\, M_{\tilde g}\, \mu\, \tan
  \beta \;I(M_{\tilde b_1}, M_{\tilde b_2}, M_{\tilde g})  \\
  I(a,b,c) &=& \frac{-1}{(a^2-b^2)(b^2-c^2)(c^2-a^2)} \left (a^2b^2 \ln
    \frac{a^2}{b^2} + b^2 c^2 \ln
    \frac{b^2}{c^2}  + c^2a^2 \ln
    \frac{c^2}{a^2}\right ). \nonumber 
\end{eqnarray} 
Moreover, the $b \bar b H_1$ coupling is dynamically generated  at $\mathcal{O}(\alpha_s)$
and can be enhanced if $\tan \beta$ is large. 
This effect can be included modifying the 
$c^\eta(bb \mathh^0)$ couplings. 
The actual effect of this modification and of the bottom mass resummation,  
Eq:~(\ref{Eq:MBRES}), is to substitute the $c^\eta(bb \mathh^0)$ couplings 
in Eq.~(\ref{eq:couplings}) as  follows 
\begin{eqnarray}
  c^\eta(bbh^0) &\to&    \frac{c^\eta(bbh^0)}{m_b} \times
  \overline{m}^{\overline{\mbox{\tiny DR}}}~
  \left (1-\frac{\Delta_b}{\tan\beta \tan \alpha} \right )  \nonumber \\
  c^\eta(bbH^0) &\to&    \frac{c^\eta(bbH^0)}{m_b} \times  \overline{m}^{\overline{\mbox{\tiny DR}}}~
  \left  (1+\frac{\Delta_b \tan \alpha}{\tan\beta} \right )  \nonumber \\
   c^\eta(bbA^0) &\to&    \frac{c^\eta(bbA^0)}{m_b} \times  \overline{m}^{\overline{\mbox{\tiny DR}}}~
 \left   (1-\frac{\Delta_b }{\tan^2 \beta} \right ).  
\end{eqnarray} 
\noindent
We have checked the cancellation of the UV divergences
among counterterms, self-energies and triangles.
This cancellation occurs separately inside 8 sectors, i.e. $s$-channel ``initial''
triangles with chirality L or R, $s$-channel ``final'' L or R,
$u$-channel up triangles (L or R) and $u$-channel down triangles (L or R).
The Box diagrams are UV finite.

\subsection{QED radiation}
\label{sub:E}

\begin{figure}
\centering
\includegraphics[width=0.7\textwidth]{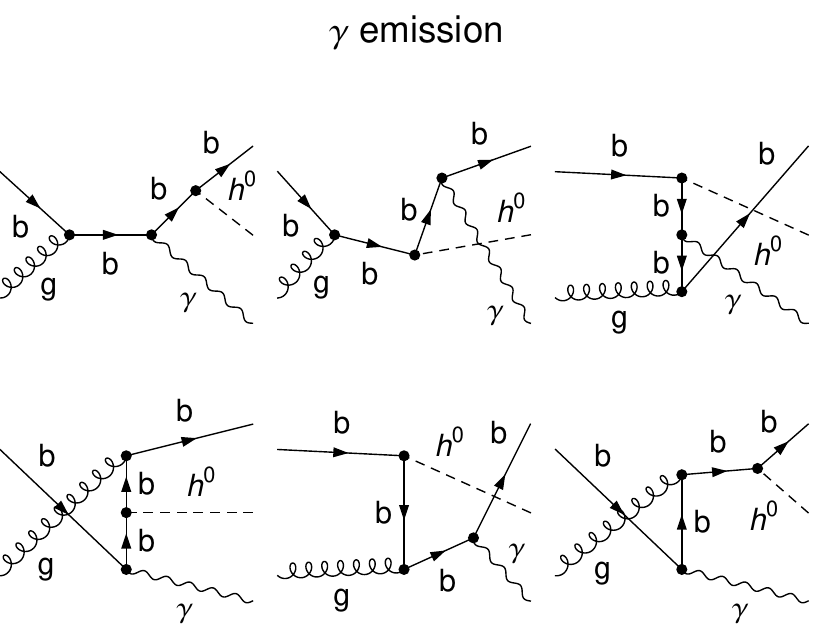}
\caption{Real photon  $\gamma$ emission.
\label{fig:qed}}
\end{figure}

The infrared (IR) singularities affecting the virtual contributions are
cancelled  including the bremsstrahlung of real photons at $\mathcal{O}(\alpha_s\alpha^2)$,
\begin{equation}
  b(p_b)~g(p_g) \to ~b(p'_b)~\mathcal{H}^0(p_{\mathcal{H}^0})~\gamma(p_\gamma) \, ,
  \label{Eq:PartonGluon_Gamma}
\end{equation}
arising from the diagrams in Figure~\ref{fig:qed}.  This contribution has been computed using 
 \verb+FeynArts+~\cite{Kublbeck:1990xc} and
\verb+FormCalc+~\cite{Hahn:1998yk}.
The integral over the photon pase space is IR divergent in the soft-photon region, {\sl i.e.}
for $p^0_\gamma \to 0$.
The IR divergences are regularized within mass 
regularization, 
giving a small mass $m_\gamma$ to the photon. The phase
space integration has been performed using the phase space slicing method.  This method
introduces a fictitious separator $\Delta E$ and 
restricts the numerical phase space integration  in the region characterized
by $p_\gamma > \Delta E$. The integral over the region $p_\gamma < \Delta E$ 
is performed analytically in the eikonal approximation~\cite{Baier:1973ms}.\\

\begin{figure}
\centering
\includegraphics[width=0.8\textwidth]{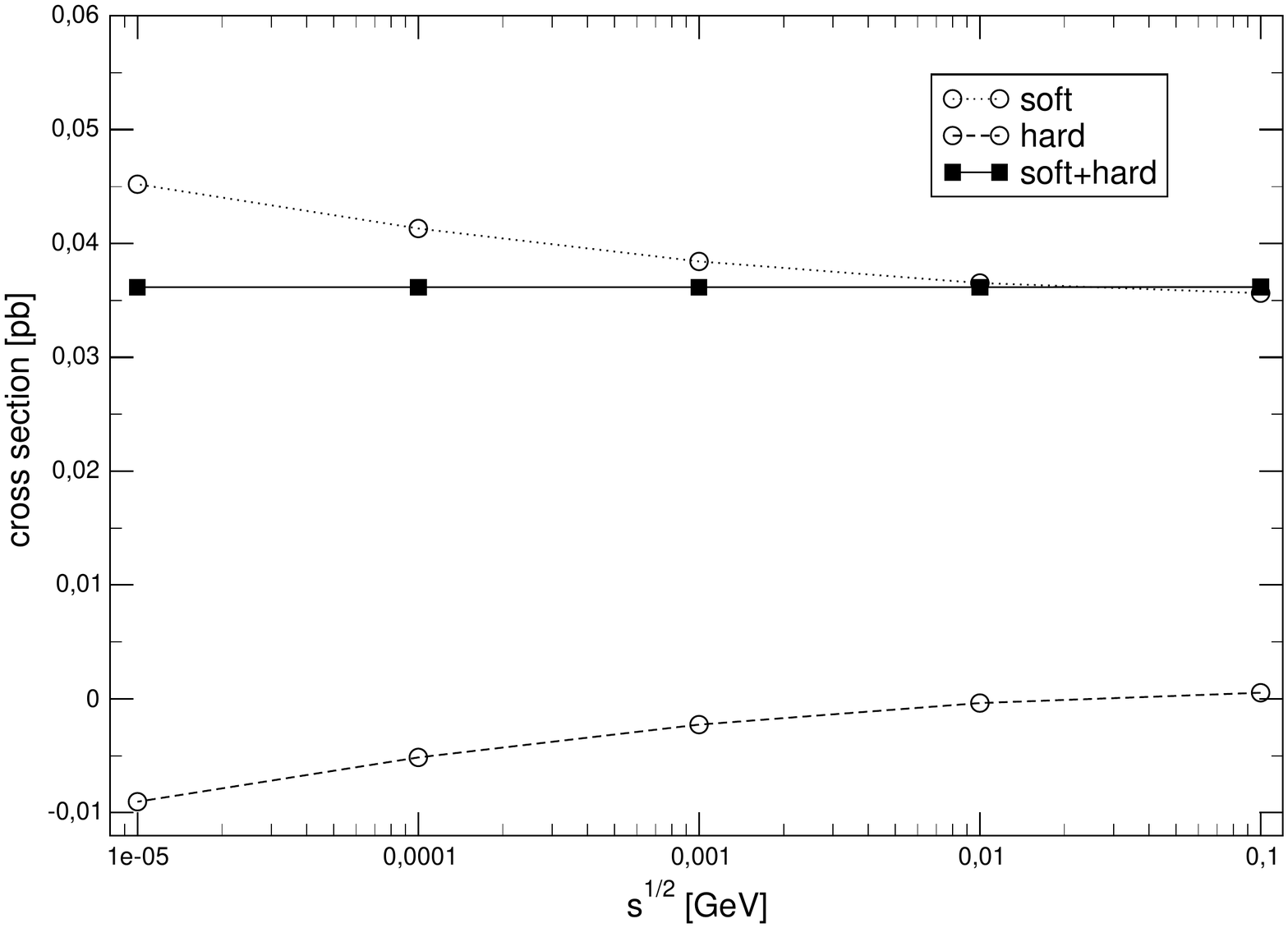}
\includegraphics[width=0.8\textwidth]{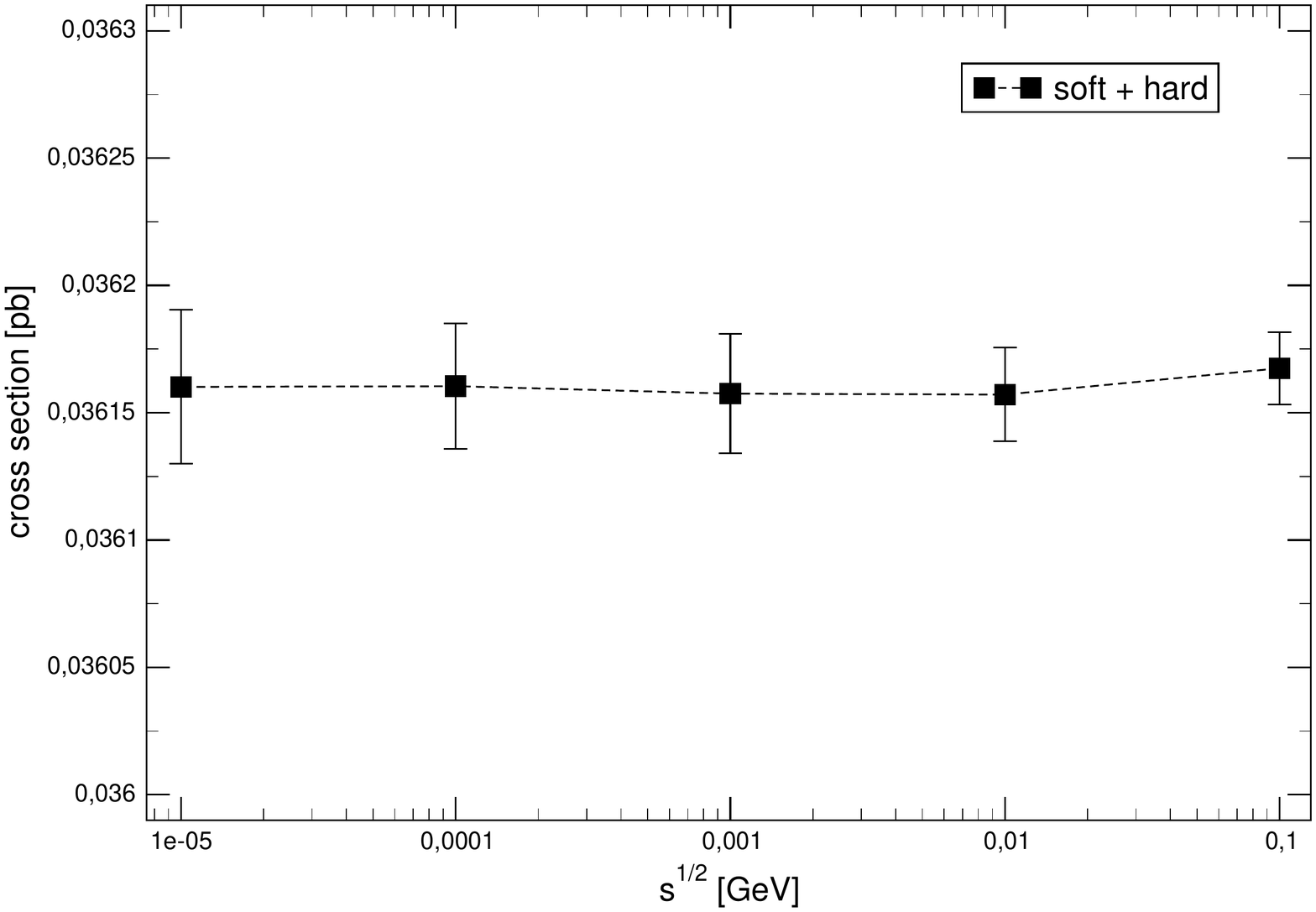}
\caption{$A^0$ production: dependence of the $\mathcal{O}(\alpha)$ soft+virtual, hard, and total
  sum corrections on the separator $\Delta E$
\label{fig:IR-A0}}
\end{figure}

\begin{figure}
\centering
\includegraphics[width=0.8\textwidth]{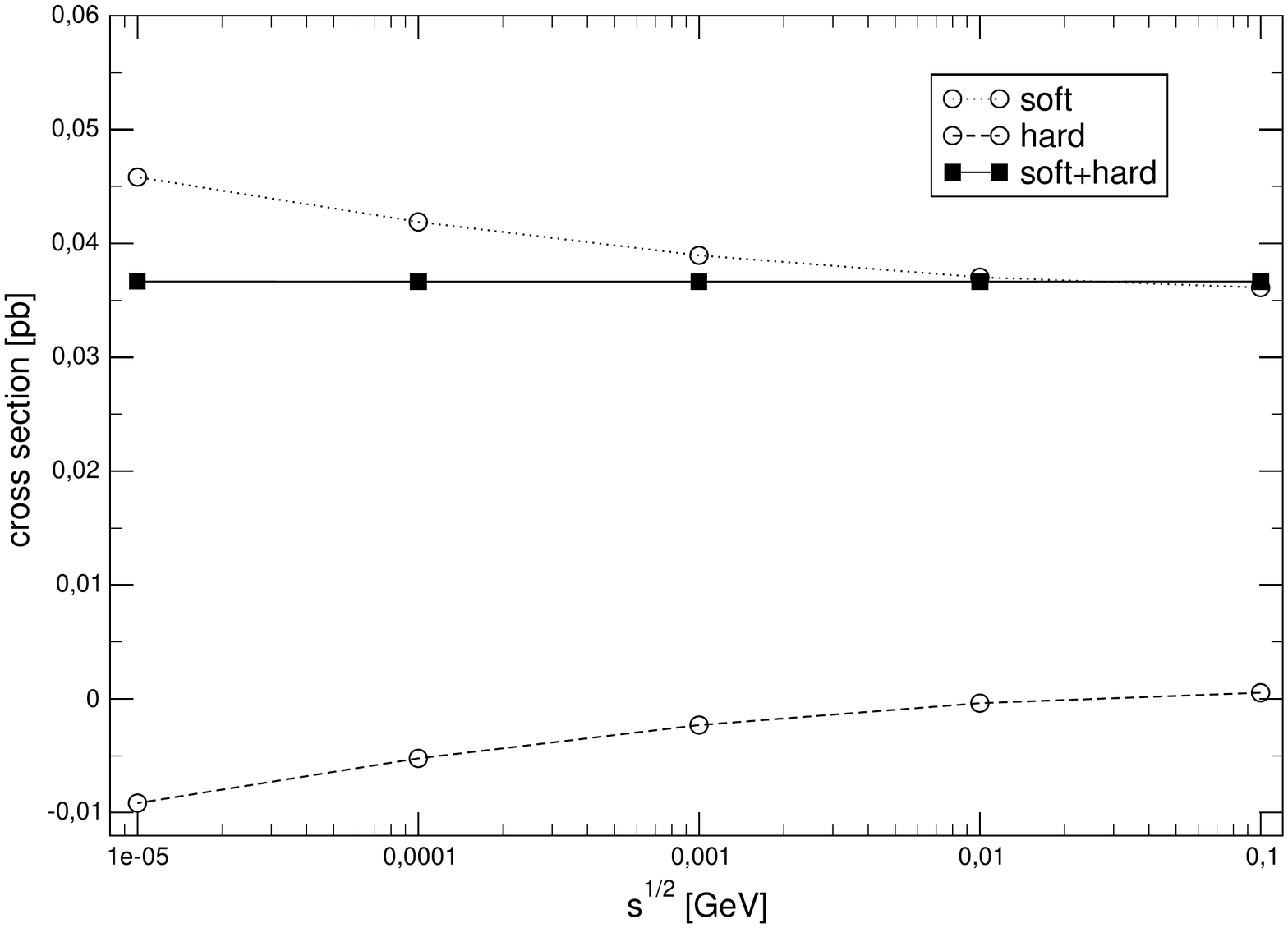}
\includegraphics[width=0.8\textwidth]{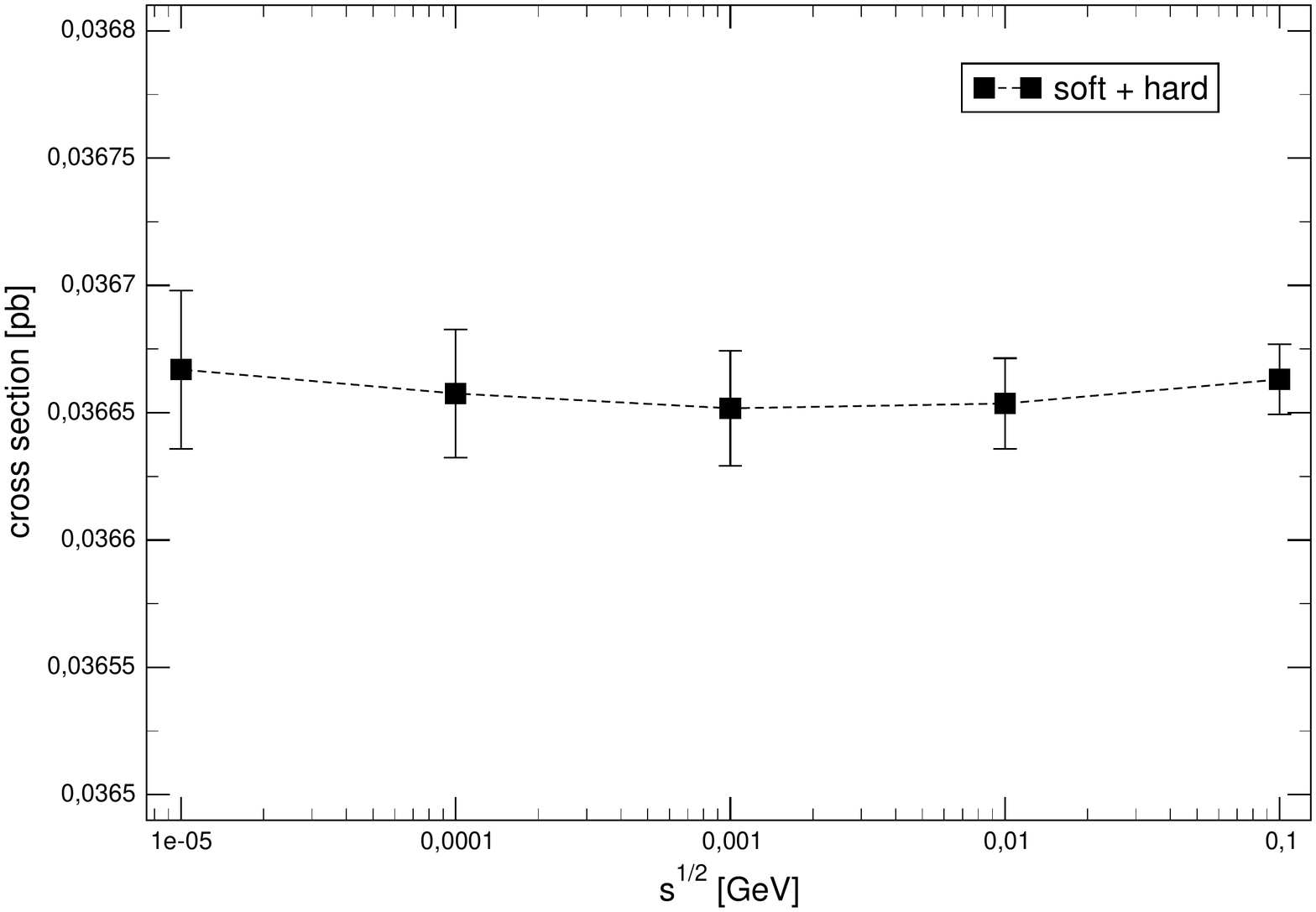}
\caption{$H^0$ production: dependence of the $\mathcal{O}(\alpha)$ soft+virtual, hard, and total
  sum corrections on the separator $\Delta E$
\label{fig:IR-H0}}
\end{figure}

\begin{figure}
\centering
\includegraphics[width=0.8\textwidth]{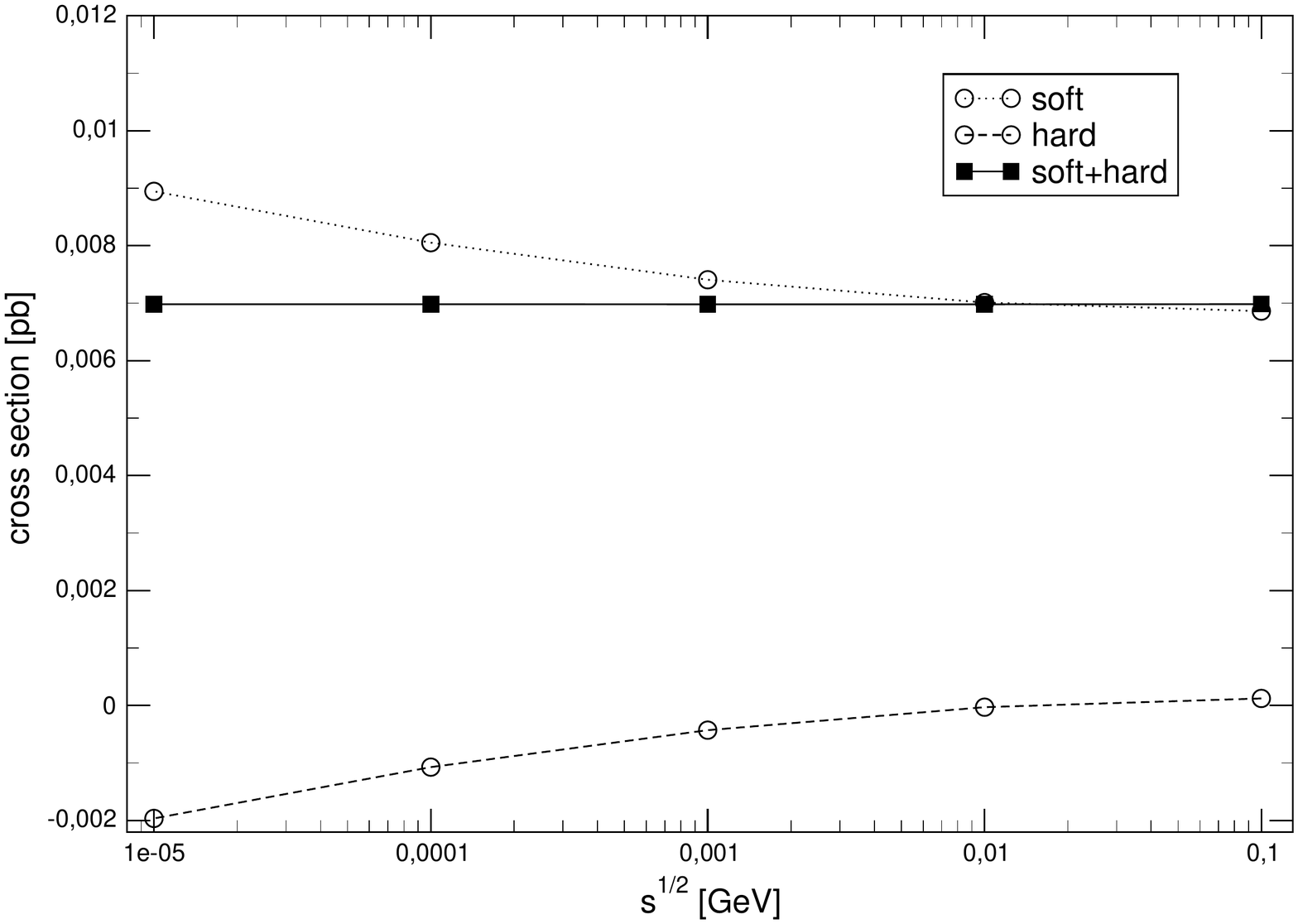}
\includegraphics[width=0.8\textwidth]{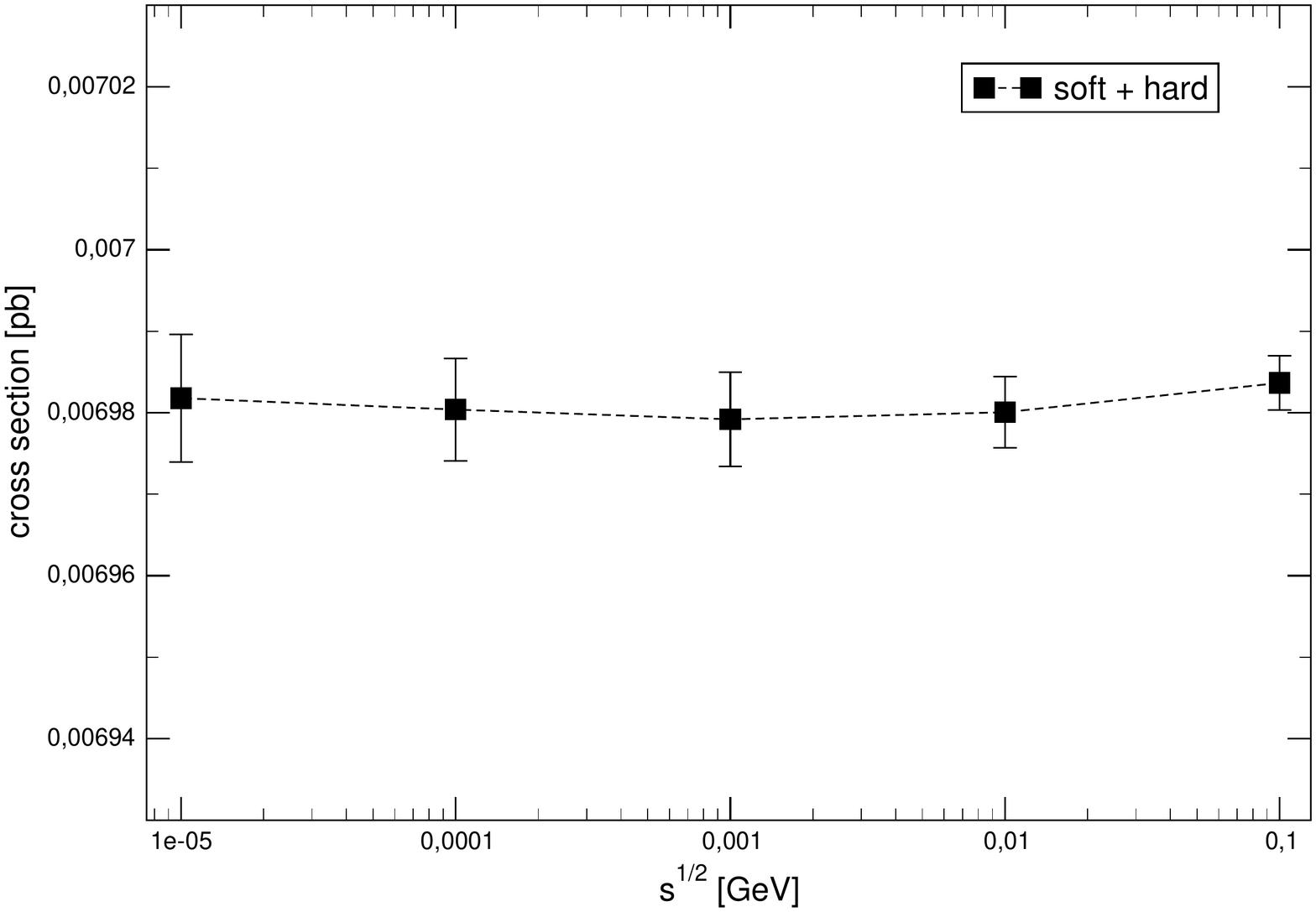}
\caption{$h^0$ production: dependence of the $\mathcal{O}(\alpha)$ soft+virtual, hard, and total
  sum corrections on the separator $\Delta E$
\label{fig:IR-h0}}
\end{figure}

Large collinear logarithms containing the bottom mass can be re-absorbed 
into the redefinition of the parton distribution function (PDF) of the bottom $f_b(x,\mu)$.
In the $\overline{\mbox{MS}}$ (DIS) factorization scheme this is achieved performing   the 
substitution~\cite{Baur:1998kt} 
\begin{eqnarray}
  f_b(x, \mu) &\to& f_b(x,\mu) \left \{ 1 - \frac{\alpha}{\pi} e^2_b\left [ 1 -\ln
      \delta_s - \ln \delta^2_s + \left ( \ln \delta_s + \frac{3}{4}
      \right ) \ln \left ( \frac{\mu^2}{m_b^2}  \right ) - \frac{1}{4}
      \lambda_{\mbox{\tiny FC}} \kappa_1  \right]  \right \} \nonumber \\
  &-& \frac{\alpha}{2 \pi} e^2_b\int_{x}^{1-\delta_s}\; \frac{dz}{z}  \;
  f_b\left( \frac{x}{z}, \mu \right ) \left [ \frac{1+z^2}{1-z} \ln \left
      (\frac{\mu^2}{m^2_b} \frac{1}{(1-z)^2} \right ) - \frac{1+z^2}{1-z}
    + \lambda_{\mbox{\tiny FC}} \kappa_2\right ],
\end{eqnarray}
and setting $\lambda_{\mbox{\tiny FC}} = 0$ ($\lambda_{\mbox{\tiny
    FC}}=1$). $\mu$ is the factorization scale, $\delta_s = 2 \Delta
E / \sqrt{s}$, while $e_b$ is the bottom charge.  $\kappa_1$ and $\kappa_2$
are defined as follows,
\begin{eqnarray}
  \kappa_1 &=& 9 +\frac{2}{3} \pi^2 +3 \ln \delta_s -2 \ln^2 \delta_s,
  \nonumber \\
  \kappa_2 &=& \frac{1+z^2}{1-z} \ln \left (\frac{1-z}{z} \right )
  -\frac{3}{2} \frac{1}{1-z} +2 z +3.
\end{eqnarray}  
We tested numerically the cancellation of IR divergences, the independence of
our results of $m_{\gamma}$ (in the sum
of the soft and virtual part) and of the separator $\Delta E$ (see Figures~\ref{fig:IR-A0},~\ref{fig:IR-H0},~\ref{fig:IR-h0}).

\subsection{Total cross sections}
\label{sub:F}

Including the finite wave function renormalization for the Higgs field we
obtain the following expressions for the tree-level differential partonic  cross section of the processes
we are considering,
\bq
d \hat \sigma^{1,1}_{bg \to b\mathcal{H}^0} = 
\frac{\beta'\, d\cos\theta}{768\,\pi\, s\, \beta}~Z_{\mathcal{H}^0}~\left |\mathcal{M}^{1/2,1/2}_{bg
      \to b\mathcal{H}^0    }\right |^2 
\eq
where $\beta=2p/\sqrt{s}$,  $\beta'=2p'/\sqrt{s}$, and $s$ is the Mandelstam variable defined in
Eq.~(\ref{Eq:Mand}); 
the NLO-EW contribution to the differential cross section reads as follows
\begin{eqnarray}
d \hat \sigma^{1,2}_{bg \to bh^0} &=&
\frac{\beta'\, d\cos\theta}{768\,\pi\, s\, \beta}~  
  Z_{h^0} \Bigg \{ \left | 1 - Z_{h^0H^0} \frac{\cos
      \alpha}{\sin \alpha}  \right |^2~\left |\mathcal{M}^{1/2,1/2}_{bg
      \to bh^0   }\right |^2 \nonumber \\
  &+&    2~\mbox{Re}
  \mathcal{M}^{1/2,1/2}_{bg \to b h^0    }\left( \mathcal{M}^{1/2,3/2}_{bg \to
      bh^0    }\right)^*  
  \Bigg \} - d \hat \sigma^{1,1}_{bg \to bh^0}, 
  \nonumber \\
d \hat \sigma^{1,2}_{bg \to bH^0} &=&
\frac{\beta'\, d\cos\theta}{768\,\pi\, s\, \beta}~ 
  Z_{H^0} \Bigg \{ \left | 1 - Z_{H^0h^0} \frac{\sin
      \alpha}{\cos \alpha}  \right |^2~\left |\mathcal{M}^{1/2,1/2}_{bg
      \to bH^0    }\right |^2 \nonumber \\
  &+&   2~\mbox{Re}
  \mathcal{M}^{1/2,1/2}_{bg \to bH^0  }\left( \mathcal{M}^{1/2,3/2}_{bg \to
      bH^0  }\right)^*  
  \Bigg \} - d \hat \sigma^{1,1}_{bg \to bH^0} ,
  \nonumber \\
d \hat \sigma^{1,2}_{bg \to bA^0} &=&
\frac{\beta'\, d\cos\theta}{768\,\pi\, s\, \beta}~
  Z_{A^0} \Bigg \{  2~\mbox{Re}
  \mathcal{M}^{1/2,1/2}_{bg \to bA^0    }\left( \mathcal{M}^{1/2,3/2}_{bg \to
      b A^0   }\right)^*  
  \Bigg \} ,
\end{eqnarray}
where the Z factors $Z_{h^0}$, $Z_{H^0}$, $Z_{A^0}$, $Z_{h^0H^0}$, and $Z_{H^0
  h^0}$  in the two renormalization schemes we are considering can be found in 
\cite{Frank:2006yh} and in \cite{Dabelstein:1994hb}.  
The partonic differential cross section for the real 
photon radiation process  reads as follows,
\bq
  d \hat \sigma^{1,2}_{bg \to b\mathcal{H}^0\gamma} =\frac{1}{4 \cdot 24} ~
  \frac{d \phi(p'_b, p_{\mathcal{H}^0},p_\gamma)}{2 \beta\, s} ~ Z_{\mathcal{H}^0}~\left |\mathcal{M}^{1/2,1}_{bg
      \to b\mathcal{H}^0   \gamma }\right |^2 
\eq
where, according to the notation introduced in~\cite{Dittmaier:1999mb}, 
$d \phi(p'_b, p_{\mathcal{H}^0},p_\gamma)$ is the three-particles phase space 
measure. The hadronic differential cross section at $\mathcal{O}(\alpha_s \alpha)$ and
$\mathcal{O}(\alpha_s \alpha^2)$ reads
\begin{eqnarray}
  d\sigma^{1,1}_{PP \to b\mathh^0}(S) &=& \int_0^1~dx_1  \int_0^1~dx_2 
  \Big[ 
  f_b(x_1, \mu) f_g(x_2, \mu)+  (x_1 \leftrightarrow x_2)
  \Big ]\nonumber \\
  &\times& d\hat \sigma^{1,1}_{bg\to b\mathh^0}(x_1 x_2 S)     \nonumber \\
  d\sigma^{1,2}_{PP \to b\mathh^0(\gamma)}(S) &=& \int_0^1~dx_1  \int_0^1~dx_2 
  \Big[ 
  f_b(x_1, \mu) f_g(x_2, \mu)+  (x_1 \leftrightarrow x_2)
  \Big ]\nonumber \\
  &\times& \Big [ d\hat \sigma^{1,2}_{bg\to b\mathh^0}(x_1 x_2 S)  +
  d\hat \sigma^{1,2}_{bg\to b\mathh^0\gamma}(x_1 x_2 S)    \Big ]
\label{eq:basicCS}
\end{eqnarray}
respectively. 
$\sqrt{S}$  is the hadronic center-of-mass energy, while  $f_i(x_i, \mu)$ is the 
parton distribution  function  
of the parton $i$ inside the proton
with a momentum fraction $x_i$ at the scale $\mu$. 
For later convenience we define the invariant mass distribution as
\begin{eqnarray}
  \frac{d\sigma^{1,1}_{PP \to b\mathh^0}}{d \sqrt{\bar s}} &=& \int_0^1~dx_1  \int_0^1~dx_2 
  \Big[ 
   f_b(x_1, \mu) f_g(x_2, \mu)+  (x_1 \leftrightarrow x_2)
  \Big ]\nonumber \\
  &\times& d\hat \sigma^{1,1}_{bg\to b\mathh^0}(x_1 x_2 S)    \delta \left (  \sqrt{x_1 x_2 S}  -\sqrt{\bar s} \right ) \nonumber \\
  \frac{d\sigma^{1,2}_{PP \to b\mathh^0}}{d \sqrt{\bar s}}&=& \int_0^1~dx_1  \int_0^1~dx_2 
  \Big[ 
  f_b(x_1, \mu) f_g(x_2, \mu)+  (x_1 \leftrightarrow x_2)
  \Big ]\nonumber \\
  &\times& \Big [ d\hat \sigma^{1,2}_{bg\to b\mathh^0}(x_1 x_2 S)  +
  d\hat \sigma^{1,2}_{bg\to b\mathh^0\gamma}(x_1 x_2 S)    \Big ]  \delta \left (  \sqrt{x_1 x_2 S}  -\sqrt{\bar s} \right )
\end{eqnarray}

\section{Numerical Results}
\label{sec:num}

The independent input parameters in the MSSM Higgs sector are the $A^0$ mass and $\tan \beta$:
since we impose the same renormalization condition for $M_{A^0}$ only $\tan
\beta$ should be converted in the change of scheme, using the one-loop relation:
\begin{equation}
  \tan \beta^{\mbox{\tiny DCPR}} = \tan \beta^{\overline{\mbox{\tiny DR}}}
  + \delta \tan \beta^{\overline{\mbox{\tiny DR}}}
  - \delta \tan \beta^{\mbox{\tiny DCPR}}, 
\end{equation}
while the OS and \drbar bottom masses $m_b^{\mbox{\tiny OS}}$ and
$m_b^{\overline{\mbox{\tiny DR}}}(\mu) $ are computed  starting from
$m_b^{\overline{\mbox{\tiny MS}}}(m_b)=4.2$ 
GeV and following the procedure described in Section 3.2.2 
of~\cite{Heinemeyer:2004xw}.\\

\begin{table}
\small{
\begin{tabular}{cc|c|c|c|c|c|c|c|c}
\hline  
& Scenario & $\;\tan \beta\;$  & $\;M_{A^0}\;$ &  $\;M_{\tilde{q},1}\;$ & $\;M_{\tilde{q},2}\;$   &  $\;M_{\tilde{q},3}\;$  & $\;M_1\;$  & $\;M_2\;$ & $\;M_{\tilde{g}}\;$ \\
\hline
& SPP$_1$       &   $15$    &  $350$   & $350$   & $350$   &  $250$  & $90$  & $150$ & $800$    \\
& SPP$_2$       & variable  &  $250$   & $500$   & $500$   &  $400$  & $90$  & $200$ & $800$    \\
\hline
\end{tabular}
\caption{Inputs parameters for the SUSY  scenarios
considered in our numerical discussion. $M_{\tilde{q}, j}$
is the common value of the breaking parameters  in the sector
of the squarks belonging to the 
$j^{\mbox{\tiny th}}$ generation.
The dimensionful parameters are given in GeV.\label{Tab:Inputs}}}
\end{table}

\begin{table}
\small{
\begin{tabular}{cc|c|c|c|c|c|c|c}
\hline  
& $\tan \beta\;$ & $\;\sigma^{\mbox{\tiny{\drbar,NLO}}}\;$  &
                   $\;\sigma^{\mbox{\tiny{\drbar,LO}}}\;$  & 
                   $\;\sigma^{\mbox{\tiny{DCPR,NLO}}}\;$  &
                   $\;\sigma^{\mbox{\tiny{DCPR,LO}}}\;$ &
                   $\;K^{\mbox{\tiny{\drbar}}}\;$ &
                   $\;K^{\mbox{\tiny{DCPR}}}\;$  & $\;$ NLO ratio \\
\hline
& 10 &  1.367    &  1.281     &    1.371    &   1.253    &    1.067 &  1.093 &   0.997\\
& 20 &  5.040    &  4.784     &    5.060    &   4.278    &    1.053 &  1.182 &   0.995\\
& 30 &  10.601   &  10.295    &    10.785   &   8.505    &    1.029 &  1.268 &   0.98\\
& 40 &  17.118   &  17.125    &    17.615   &   13.038   &    0.999 &  1.350 &  0.97\\
\hline
\end{tabular}
\caption{$A^0$ production, SPP$_2$ spectra: total cross sections [pb], $K$-factors and NLO
  \drbar/DCPR ratio}\label{tab:A0}}
\end{table}

\begin{table}
\small{
\begin{tabular}{cc|c|c|c|c|c|c|c}
\hline  
& $\tan \beta\;$ & $\;\sigma^{\mbox{\tiny{\drbar,NLO}}}\;$  &
                   $\;\sigma^{\mbox{\tiny{\drbar,LO}}}\;$  & 
                   $\;\sigma^{\mbox{\tiny{DCPR,NLO}}}\;$  &
                   $\;\sigma^{\mbox{\tiny{DCPR,LO}}}\;$ &
                   $\;K^{\mbox{\tiny{\drbar}}}\;$ &
                   $\;K^{\mbox{\tiny{DCPR}}}\;$  & $\;$ NLO ratio \\
\hline
&10 &  1.338 &  1.260 &  1.340 &  1.234 &  1.061 &  1.086 &  0.998\\
&20 &  5.133 &  4.857 &  5.099 &  4.334 &  1.056 &  1.176 &  1.006\\
&30 &  10.975 &  10.461 &  10.715 &  8.488 &  1.049 &  1.262 &  1.024\\
&40 &  18.613 &  17.918 &  17.811 &  13.248&  1.038 &  1.344 &  1.045\\
\hline
\end{tabular}
\caption{$H^0$ production SPP$_2$ spectra: total cross sections [pb], $K$-factors and NLO
  \drbar/DCPR ratio}\label{tab:H0}}
\end{table}

\begin{table}
\small{
\begin{tabular}{cc|c|c|c|c|c|c|c}
\hline  
& $\tan \beta\;$ & $\;\sigma^{\mbox{\tiny{\drbar,NLO}}}\;$  &
                   $\;\sigma^{\mbox{\tiny{\drbar,LO}}}\;$  & 
                   $\;\sigma^{\mbox{\tiny{DCPR,NLO}}}\;$  &
                   $\;\sigma^{\mbox{\tiny{DCPR,LO}}}\;$ &
                   $\;K^{\mbox{\tiny{\drbar}}}\;$ &
                   $\;K^{\mbox{\tiny{DCPR}}}\;$  & $\;$ NLO ratio \\
\hline
&10 &  0.282 &  0.248 &  0.282 &  0.243 & 1.135 &  1.156 &  1.002\\
&20 &  0.255 &  0.254 &  0.254 &  0.230 & 1.005 &  1.107 &  1.003\\
&30 &  0.228 &  0.258 &  0.230 &  0.217 & 0.882 &  1.059 &  0.988\\
&40 &  0.204 &  0.267 &  0.213 &  0.211 & 0.764 &  1.012 &  0.955\\
\hline
\end{tabular}
\caption{$h^0$ production SPP$_2$ spectra: total cross sections [pb], $K$-factors and NLO
  \drbar/DCPR ratio}\label{tab:h0}}
\end{table}

For the numerical evaluations we used the supersymmetric scenario  SPP$_{1}$ and a 
class of points of the parameter space SPP$_2$, with variable $\tan \beta =10,20,30,40$. The input parameters 
characterizing these scenarios are summarized in Table~\ref{Tab:Inputs}.
The sparticle masses and mixing angles have been obtained with the code \verb+FeynHiggs+ \cite{Heinemeyer:1998yj}.  
The one-loop Higgs masses are numerically computed by finding the zero of
inverse one-loop propagator matrix determinant
\begin{equation}
  [ k^2 - M_{H^0}^2 + \hat{\Sigma}_{H^0}(k^2)] [ k^2 - M_{h^0}^2 +
  \hat{\Sigma}_{h^0}(k^2)] - \hat{\Sigma}^2_{H^0h^0}(k^2)=0. 
\end{equation} 
Since we require semi-inclusive production (i.e. the bottom quark
must be tagged) we impose the
following kinematical cuts on the bottom in the final state,
limiting the transferred momentum $p_{b,T}>20$ GeV (due to resolution
limitations of the hadronic calorimeter) and the rapidity
$|y_b|<2$ (in order to be able to perform inner tracking). 
The process we are considering is leading order in QCD. 
Therefore, analogously to ~\cite{WjetProd, Beccaria:2008jq,  Beccaria:2009my}, we use 
a   LO QCD PDF set, namely the
LO CTEQ6L~\cite{Pumplin:2002vw}.   Our choice is justified since the QED effects in the DGLAP evolution equations are known to be small~\cite{Roth:2004ti}.
The
factorization of the bottom PDF has been performed
in the DIS scheme, with factorization scale 
$\mu = M_{\mathh^0}+m^{\mbox{\tiny OS}}_b$.\\
 
\begin{figure}
\centering
\includegraphics[width=1.0\textwidth]{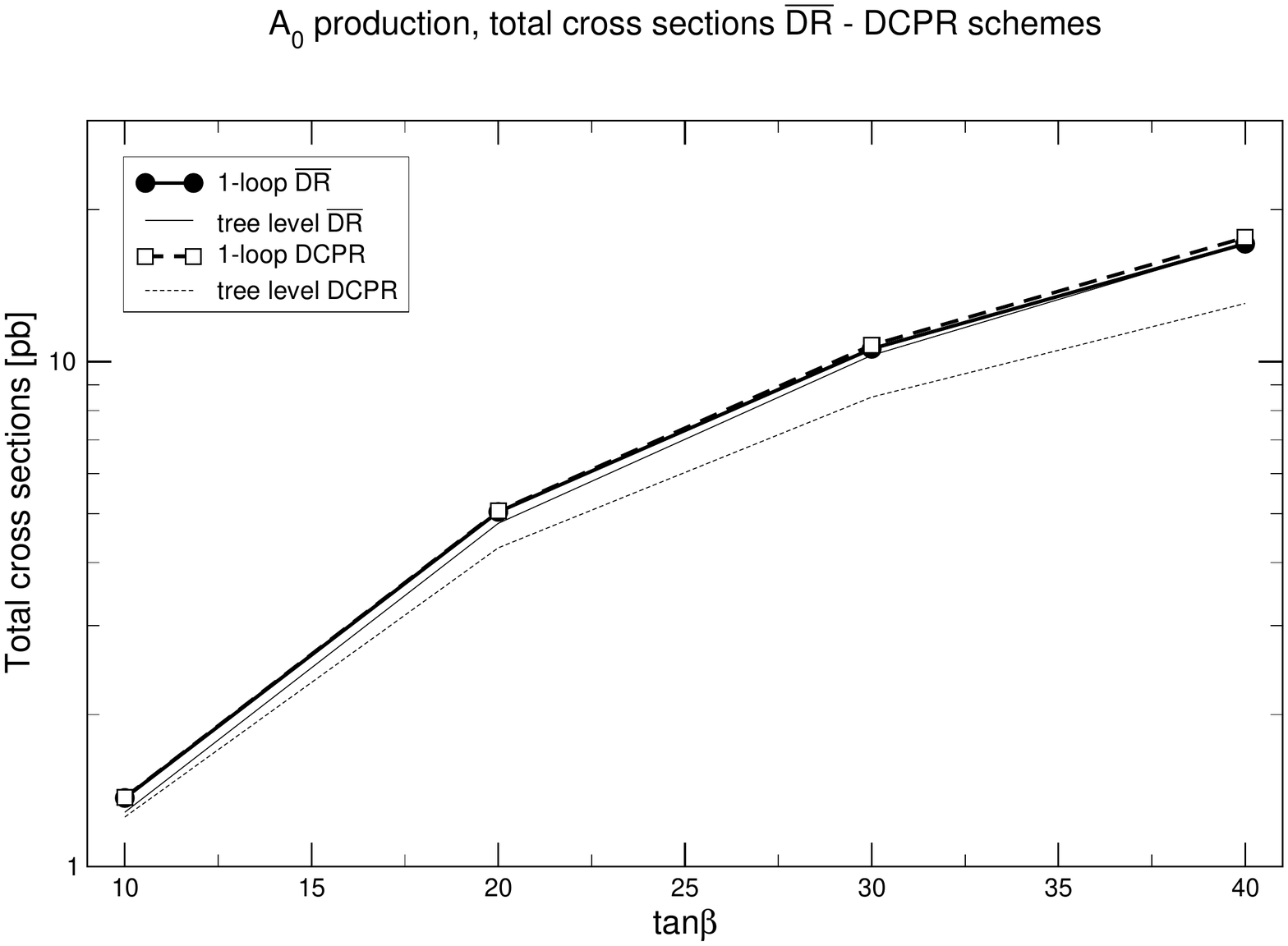}
\caption{Total LO and NLO cross sections in the \drbar and DCPR schemes, $A^0$
  production; $M_{A^0}=250$ GeV, $p_{b,T}>20$ GeV, $|y_b|<2$.
\label{fig:1}}
\end{figure}

\begin{figure}
\centering
\includegraphics[width=1.0\textwidth]{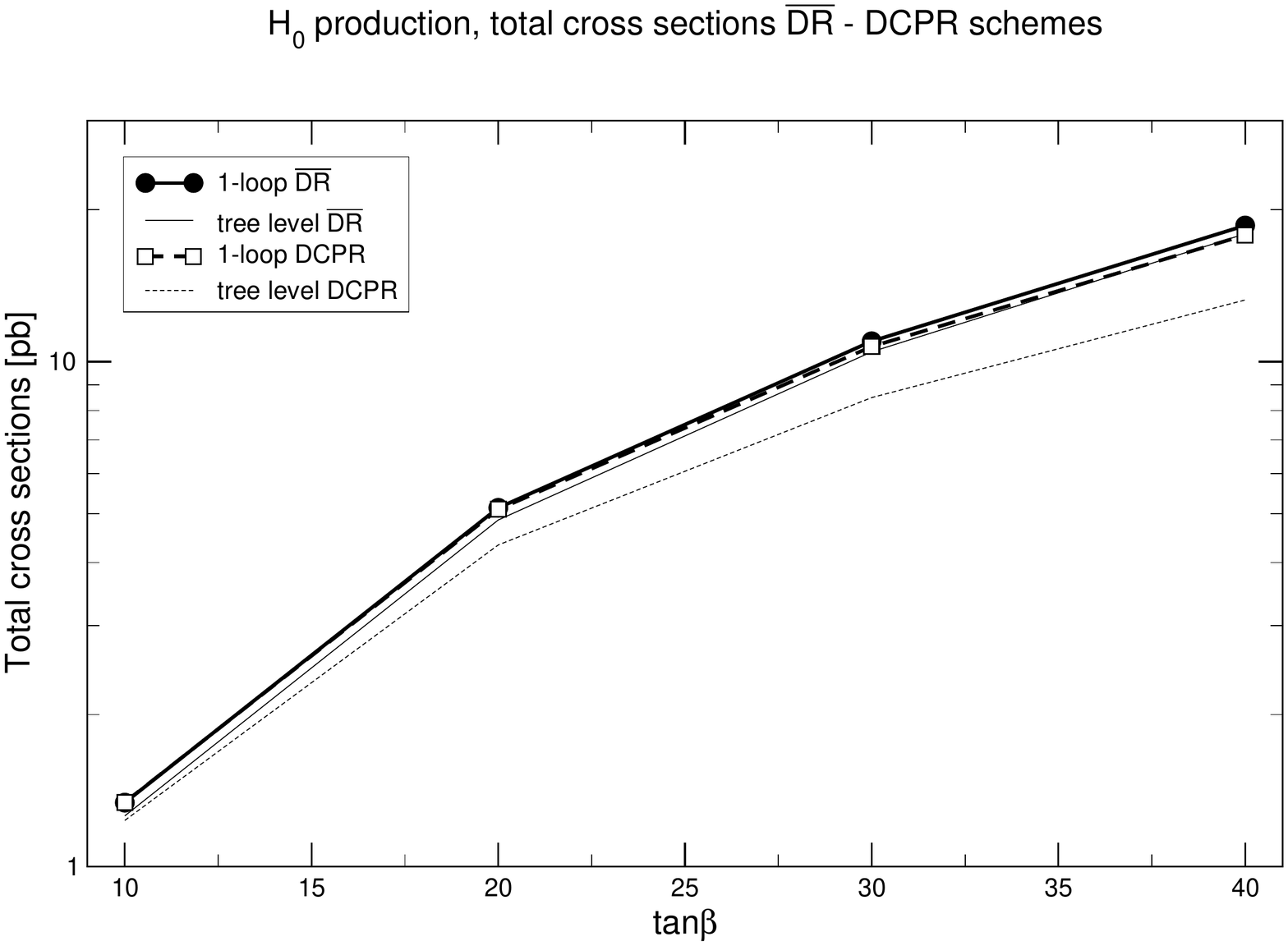}
\caption{Total LO and NLO cross sections in the \drbar and DCPR schemes, $H^0$
  production; $M_{A^0}=250$ GeV, $p_{b,T}>20$ GeV, $|y_b|<2$.
\label{fig:2}}
\end{figure}

\begin{figure}
\centering
\includegraphics[width=1.0\textwidth]{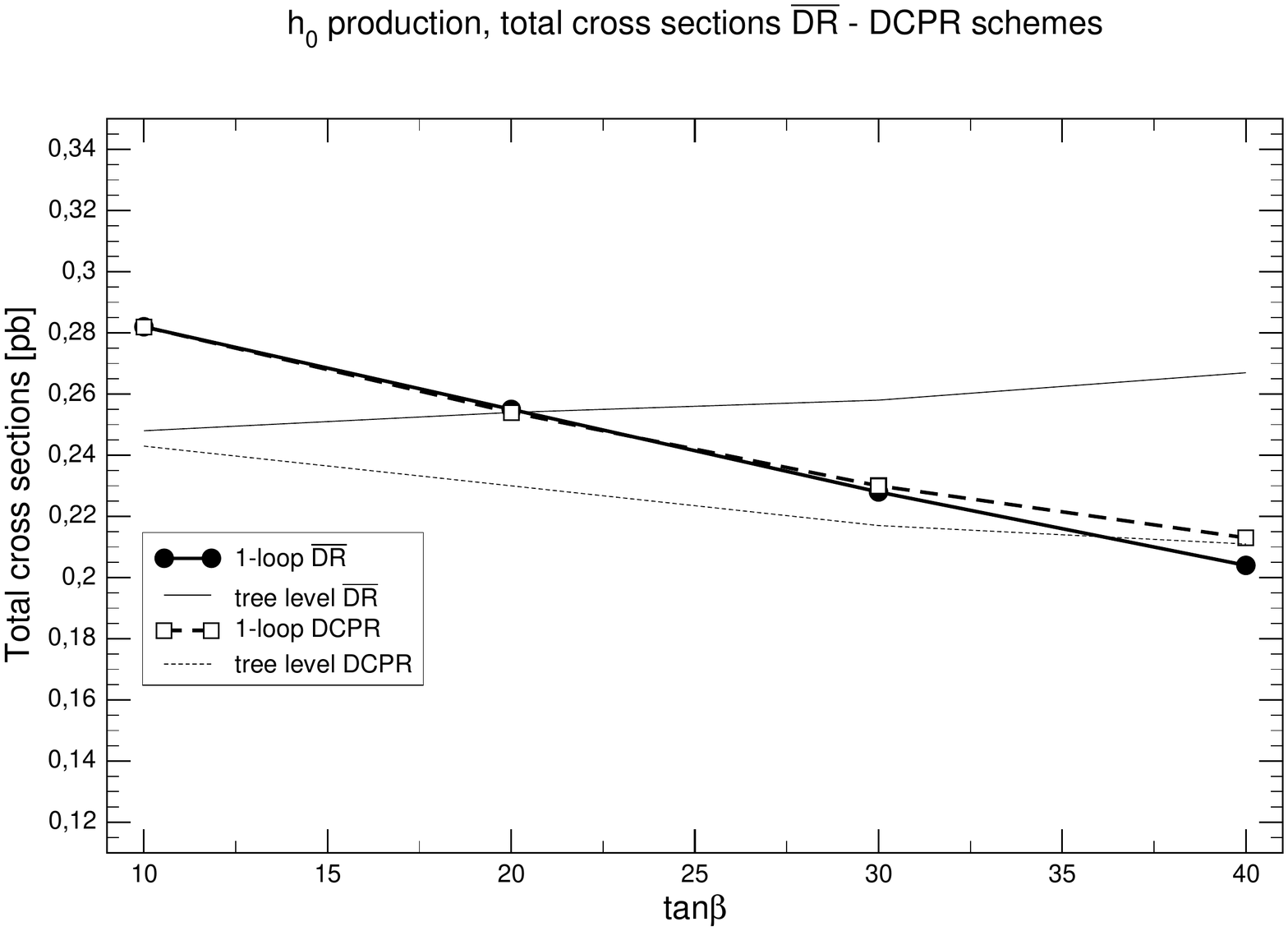}
\caption{Total LO and NLO cross sections in the \drbar and DCPR schemes, $h^0$
  production; $M_{A^0}=250$ GeV, $p_{b,T}>20$ GeV, $|y_b|<2$.
\label{fig:3}}
\end{figure}

In Figures~\ref{fig:1},~\ref{fig:2},~\ref{fig:3} we show the total cross section for $A^0,\,H^0$ and $h^0$
production in the class of supersymmetric scenarios SPP$_2$, 
as functions of $\tan \beta$. We  present both the results 
in the  \drbar and in the  DCPR schemes.
The numerical values and the $K$-factors in the two schemes (defined as usual as
the ratios $\sigma^{NLO}/\sigma^{LO}$; note that the LO is computed with the
resummed/modified SUSY QCD coupling, so our $K$-factors account of the pure
electroweak NLO effect), as well as the
ratios of the NLO cross sections in the two scheme 
are reported in Table~\ref{tab:A0},~\ref{tab:H0},~\ref{tab:h0}.\\

As one sees, the values of the
total cross sections do coincide in the overall range, apart from small differences of
the few percent size 
for very large $\tan \beta$ values. 
This confirms our expectation that at the NLO level the two schemes should be
equivalent, and also provides an important check of the reliability of our
calculations.\\

Having verified the realistic one-loop equivalence of the
two schemes, we have decided to perform our analysis in the \drbar scheme. The
main  theoretical reasons of our choice have been fully illustrated
in~\cite{schemes}. 
In particular this scheme is known to be generally more stable
numerically: our results confirm mainly this expectation but it is worth to
note that for $h^0$ production both schemes can produce (in different $\tan
\beta$ regions) relatively large effects; nevertheless the good agreement
between the two schemes leads to suppose that the perturbative expansion is
well behaved, and NNLO effects are well under control.\\

\begin{figure}
\centering
\includegraphics[width=1.0\textwidth]{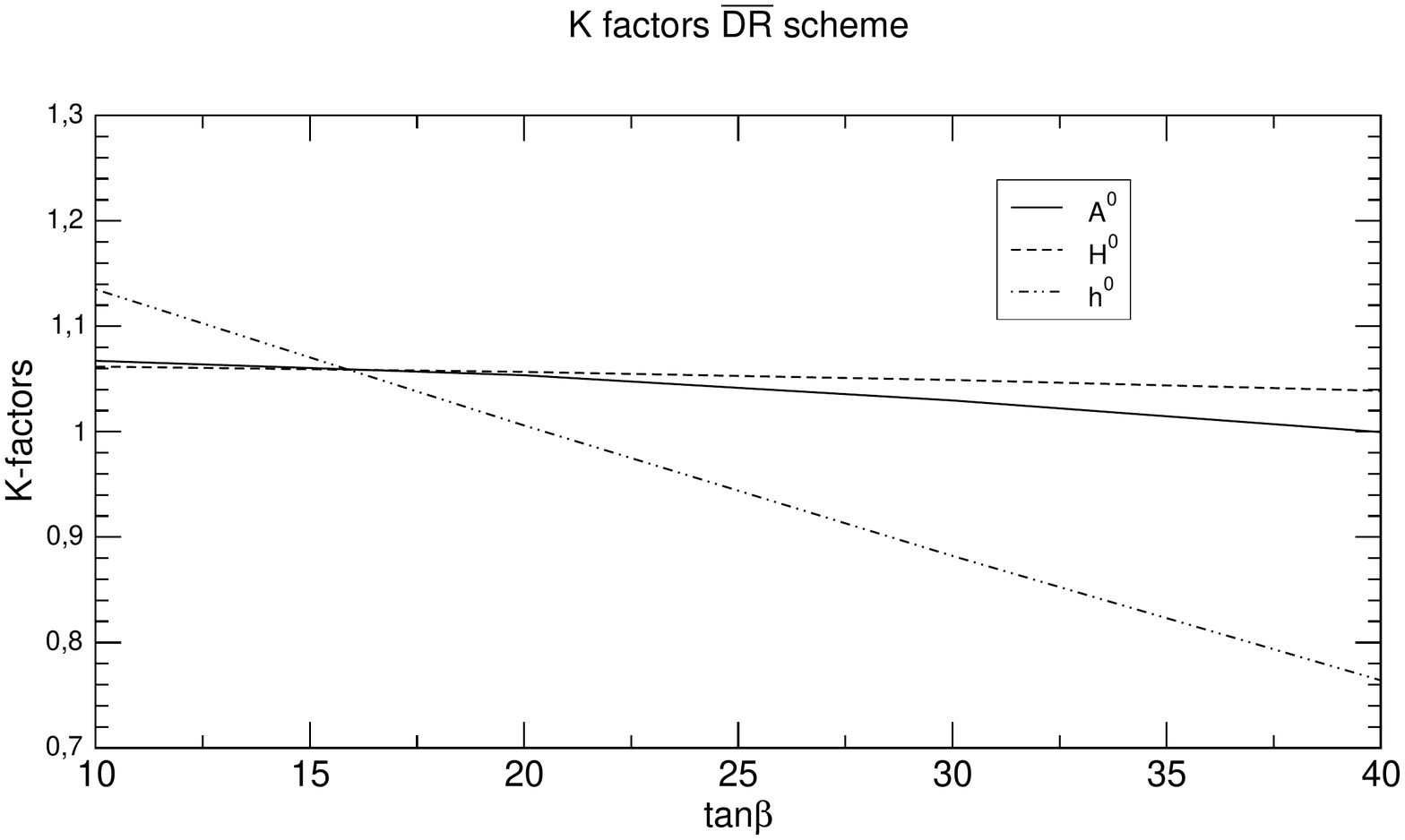}
\caption{$K$-factors for $A^0,\,H^0$ and $h^0$ production, \drbar scheme.  
$M_{A^0}=250$ GeV, $p_{b,T}>20$ GeV, $|y_b|<2$.
\label{fig:K}}
\end{figure}

\begin{figure}
\centering
\includegraphics[width=1.0\textwidth]{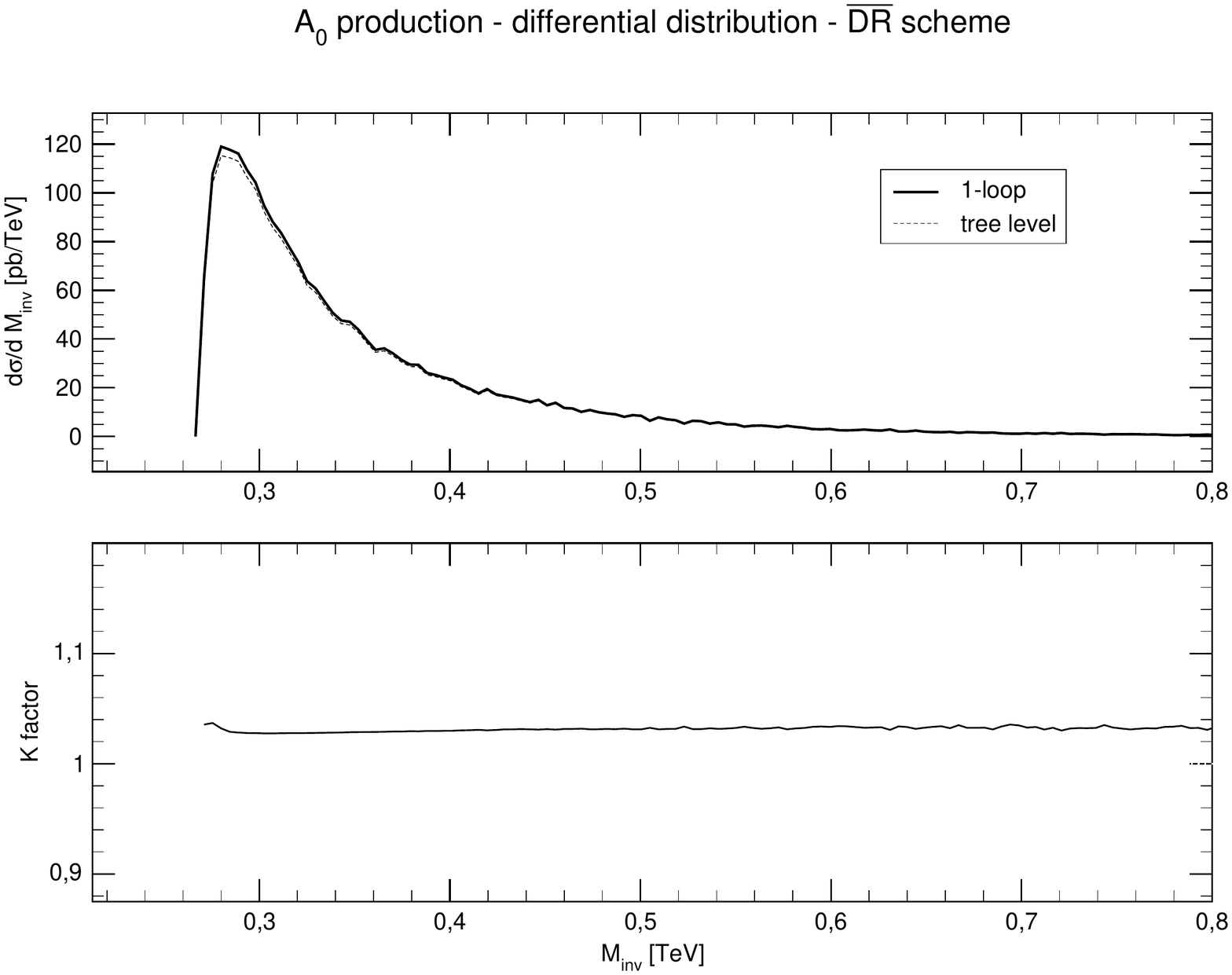}
\caption{Invariant mass distribution, $A^0$ production, \drbar scheme.  
$M_{A^0}=250$ GeV, $\tan \beta=30$ $p_{b,T}>20$ GeV, $|y_b|<2$.
\label{fig:A0-DR-DCPR-distr}}
\end{figure}

\begin{figure}
\centering
\includegraphics[width=1.0\textwidth]{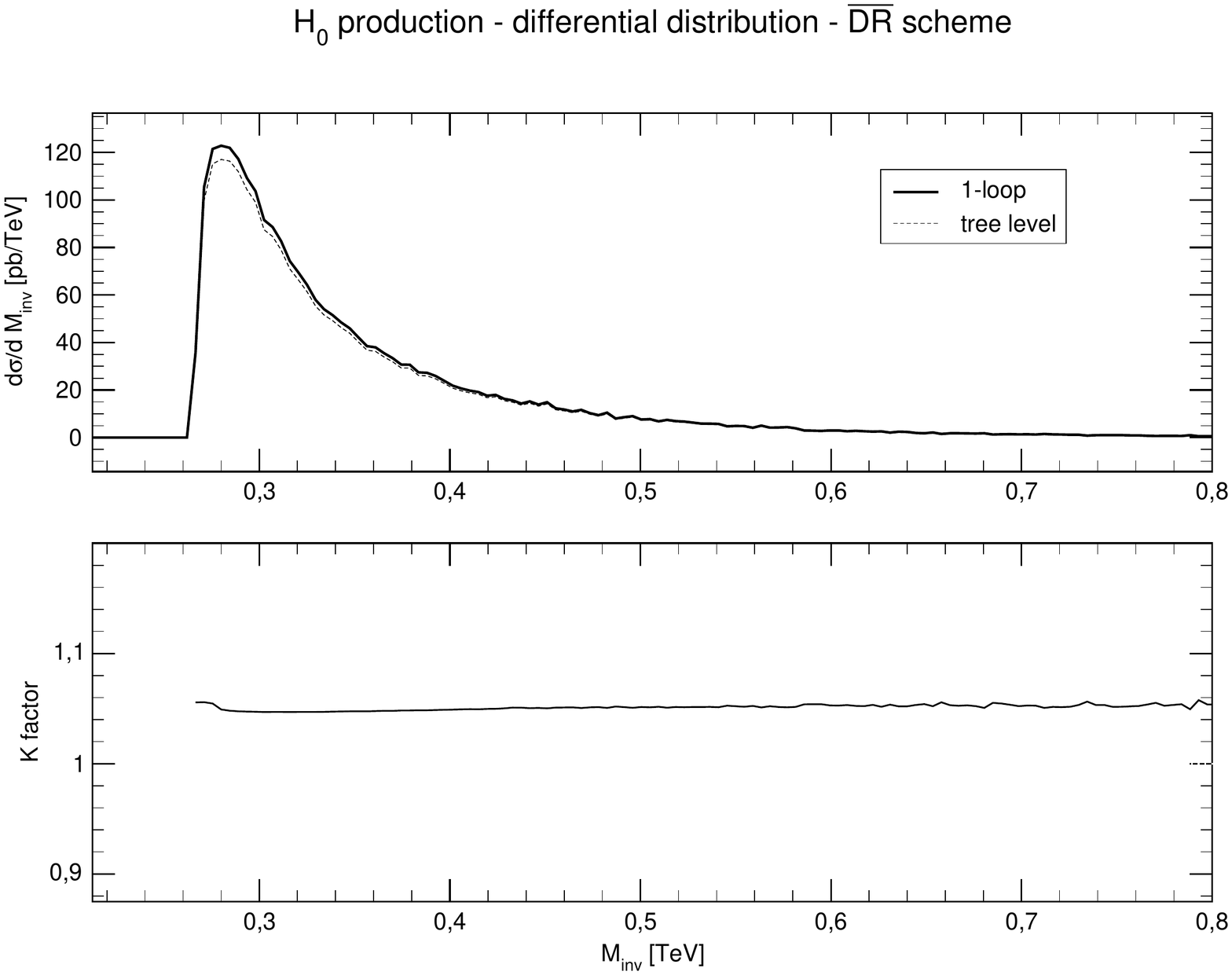}
\caption{Invariant mass distribution, $H^0$ production, \drbar scheme.  
$M_{A^0}=250$ GeV, $\tan \beta=30$ $p_{b,T}>20$ GeV, $|y_b|<2$.
\label{fig:H0-DR-DCPR-distr}}
\end{figure}

\begin{figure}
\centering
\includegraphics[width=1.0\textwidth]{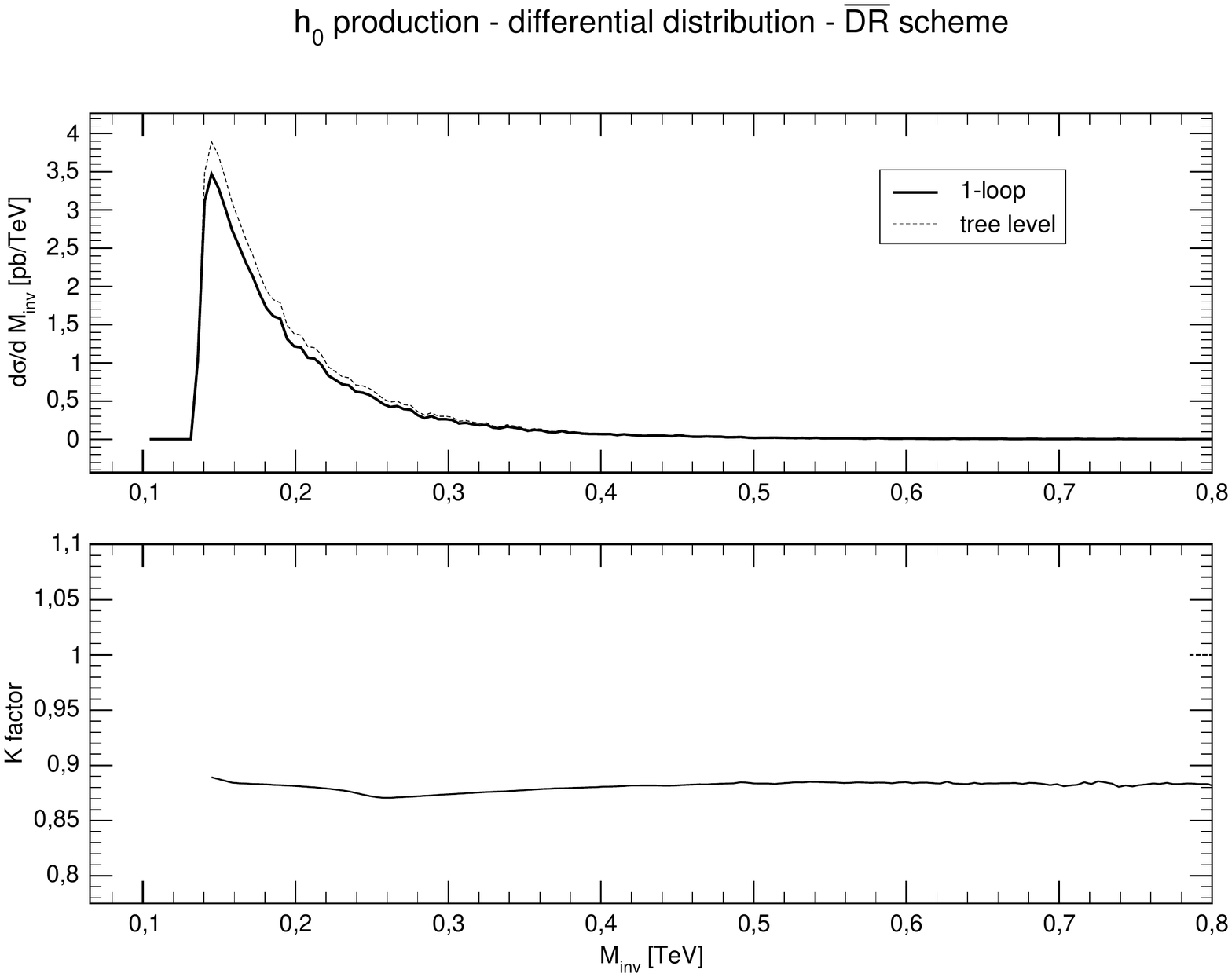}
\caption{Invariant mass distribution, $h^0$ production, \drbar scheme.  
$M_{A^0}=250$ GeV, $\tan \beta=30$ $p_{b,T}>20$ GeV, $|y_b|<2$.
\label{fig:h0-DR-DCPR-distr}}
\end{figure}

\begin{figure}
\centering
\includegraphics[width=1.0\textwidth]{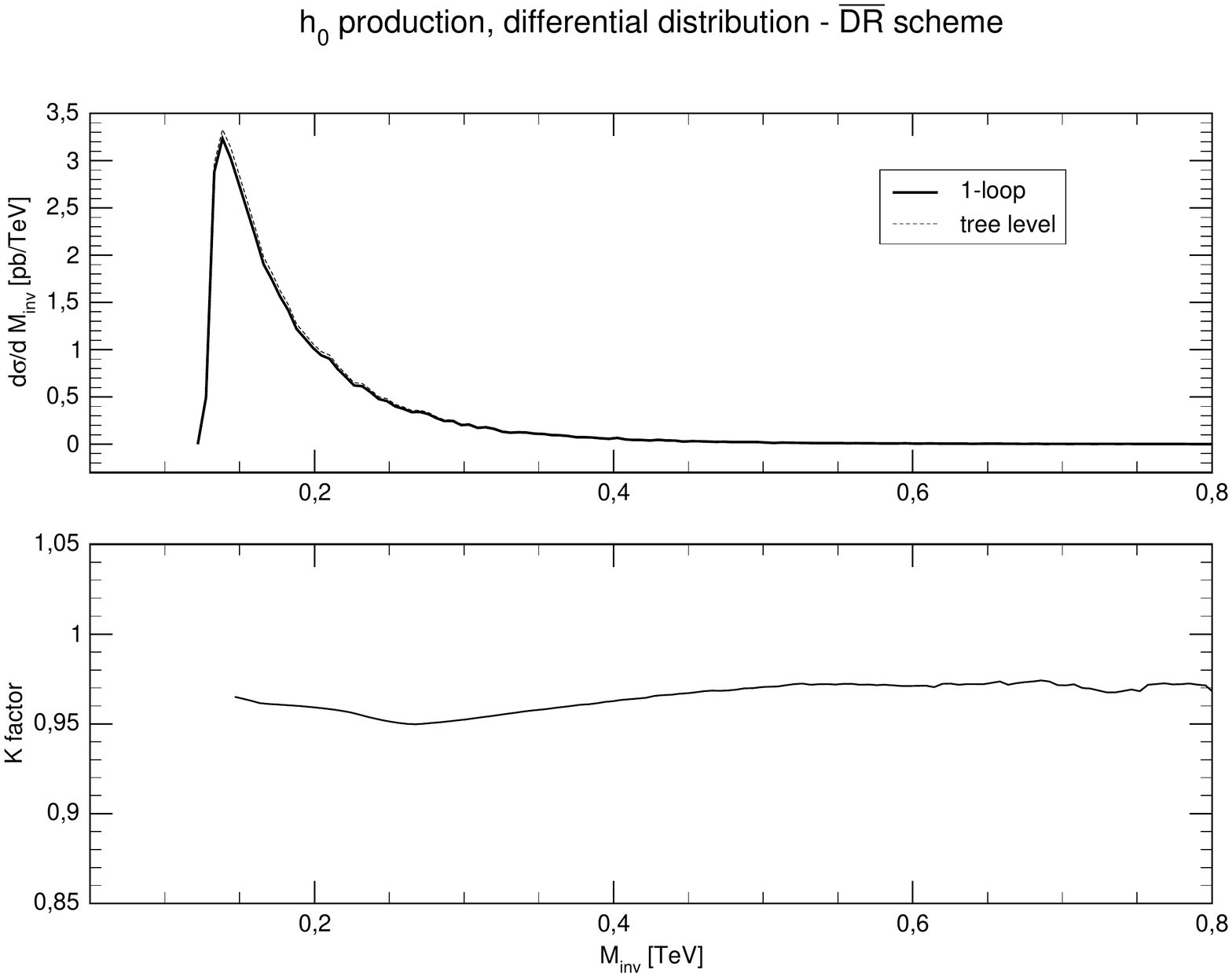}
\caption{Invariant mass distribution, $h^0$ production, \drbar scheme.  
$M_{A^0}=350$ GeV, $\tan \beta=15$ $p_{b,T}>20$ GeV, $|y_b|<2$.
\label{h0-MA350-tb15-K-QED}}
\end{figure}

Figure~\ref{fig:K} shows the $K$-factors for the three Higgs bosons in \drbar as function of $\tan \beta$   
while Figures~\ref{fig:A0-DR-DCPR-distr},~\ref{fig:H0-DR-DCPR-distr},~\ref{fig:h0-DR-DCPR-distr} show, 
for the scenario SPP$_2$ $\tan \beta = 30$, the invariant mass distribution  and the
relative NLO effect. In the next Figures~\ref{A0-MA350-tb15-K-QED},~\ref{H0-MA350-tb15-K-QED},~\ref{h0-MA350-tb15-K-QED} 
we again plot the differential distributions for the SPP$_1$ scenario; the total cross sections for this scenario are reported in
Table~\ref{tab:tb15}.\\

\begin{table}
\small{
\begin{tabular}{cc|c|c|c}
\hline  
& $\mathh^0\;$   & $\;\sigma^{\mbox{\tiny{\drbar,NLO}}}\;$  &
                   $\;\sigma^{\mbox{\tiny{\drbar,LO}}}\;$  & 
                   $\;K^{\mbox{\tiny{\drbar}}}\;$ \\
                 
\hline
&$A^0$ &  0.768 &  0.724 &  1.060 \\
&$H^0$ &  0.769 &  0.727 &  1.056 \\
&$h^0$ &  0.213 &  0.222 &  0.961 \\
\hline
\end{tabular}
\caption{SPP$_1$ spectrum: total cross sections [pb] for the three Higgs and
  \drbar  $K$-factors}\label{tab:tb15}}
\end{table}

From  inspection of the figures, one can
draw the following main conclusions:
\begin{enumerate}
\item The $K$-factors for $H^0,A^0$ are systematically small for large $\tan \beta$, and
  would reach a larger size (roughly, 8 \%) for small $\tan \beta$
  values around 10. 

\item The $K$-factor for $h^0$ varies drastically with $\tan \beta$, changing from positive
  values of about 15 \% for $\tan \beta$ around 10 to negative values of
  about 25 \%  for $\tan \beta$ around 40. These extreme negative
  and positive values are of a size that cannot be ignored in a dedicated
  experimental analysis.
\end{enumerate}
These features follow from the THDM structure and the $h^0-H^0$ mixing where
$\alpha$ is close to $\beta-\pi/2$ leading to a $\tan\beta$ enhancement in the $h^0$ case
but to a $1/\tan\beta$ suppression in the $H^0,A^0$ cases.\\

This, we believe, is the main message of our calculation: 
while for sure the QCD NLO are the dominant corrections (of order  $20-40
\%$ depending on the Higgs mass, see for example~\cite{SI-NLOQCD1}),
as it was to be  expected from the analysis of Dittmaier
et al.~\cite{I-EW}, the one-loop electroweak contribution in the semi-inclusive
bottom-Higgs production processes  must not be a priori considered as negligible.

\section{Numerical Approximations}
\label{sec:IBA}

Having performed the calculation of complete one-loop effect on the process, we shall consider
the possibility of 
simpler, effective approximations to the full
and long calculation, that may be used to obtain a quicker and qualitative
description of the results.\\

With this purpose we have first considered 
the ``improved Born Approximation''
(IBA) following  the prescriptions given in~\cite{I-EW}: the IBA is obtained is
this case by including in the definition of $\Delta_b$ (see eq.~\ref{Eq:MBRES}) the electroweak
contributions and then replacing the mixing angle $\alpha$ with the effective value $\alpha_{eff}$, obtained 
by the diagonalization of the one loop mass matrix 
\begin{eqnarray}
\left(
\begin{array}{cc}
m_{h^0}^2-\hat{\Sigma}_{h^0}(m_{h^0}^2) & -\hat{\Sigma}_{h^0H^0}(\frac{1}{2}(m_{h^0}^2 + m_{H^0}^2)) \\
 -\hat{\Sigma}_{h^0H^0}(\frac{1}{2}(m_{h^0}^2 + m_{H^0}^2)) & m_{H^0}^2-\hat{\Sigma}_{H^0}(m_{H^0}^2)
\end{array}\right)
\end{eqnarray}
The effect of the latter redefinition of $\alpha$ is negligible for
$H^0$ and $A^0$, but significant for $h^0$.\\ 

\begin{figure}
\centering
\includegraphics[width=1.0\textwidth]{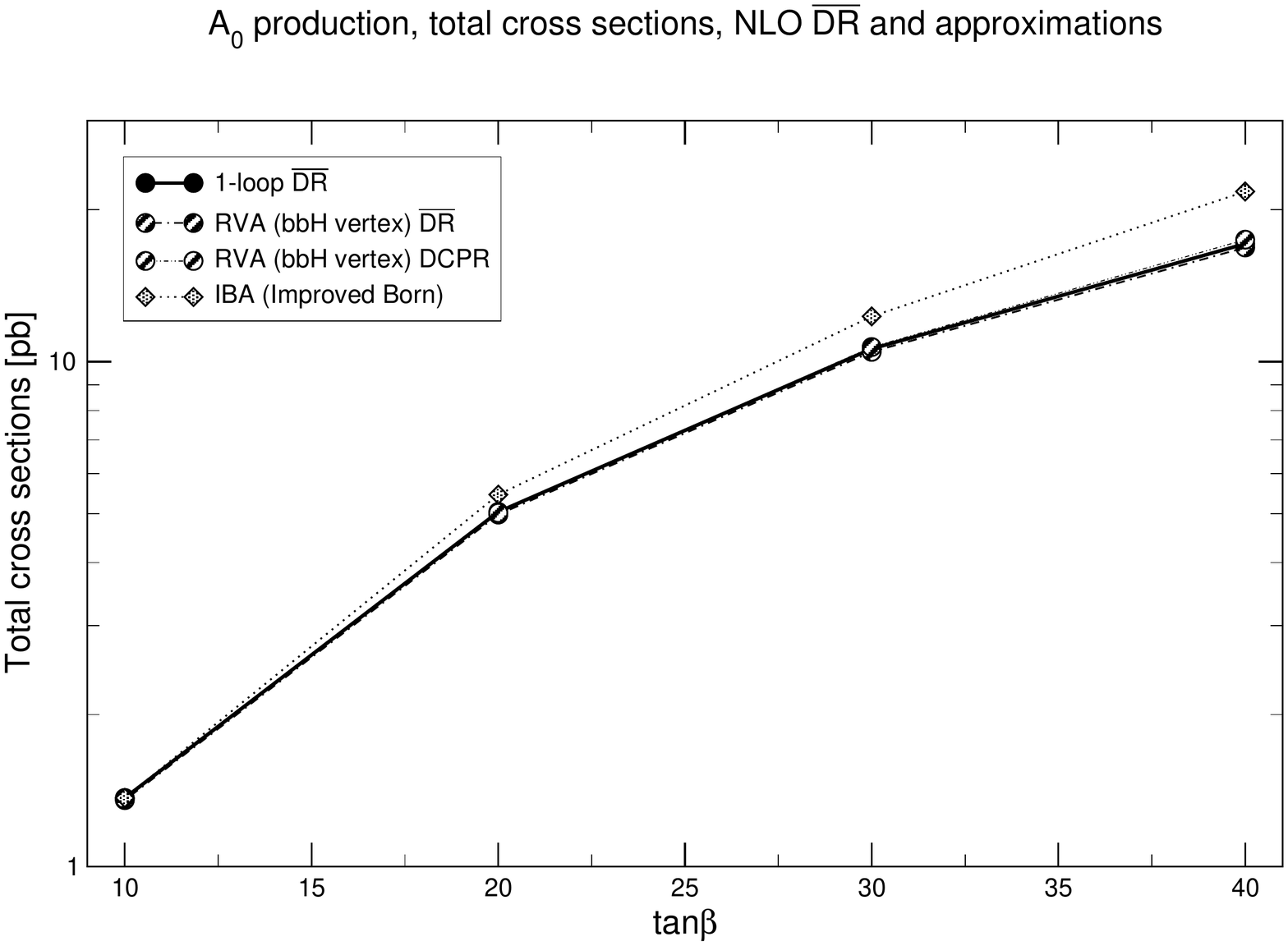}
\caption{Comparison of the total NLO cross sections: NLO \drbar, RVA (\drbar
  and DCPR) and Improved Born Approximantion (IBA), $A^0$
  production; $M_{A^0}=250$ GeV, $p_{b,T}>20$ GeV, $|y_b|<2$.
\label{fig:1IBA}}
\end{figure}
\begin{figure}
\centering
\includegraphics[width=1.0\textwidth]{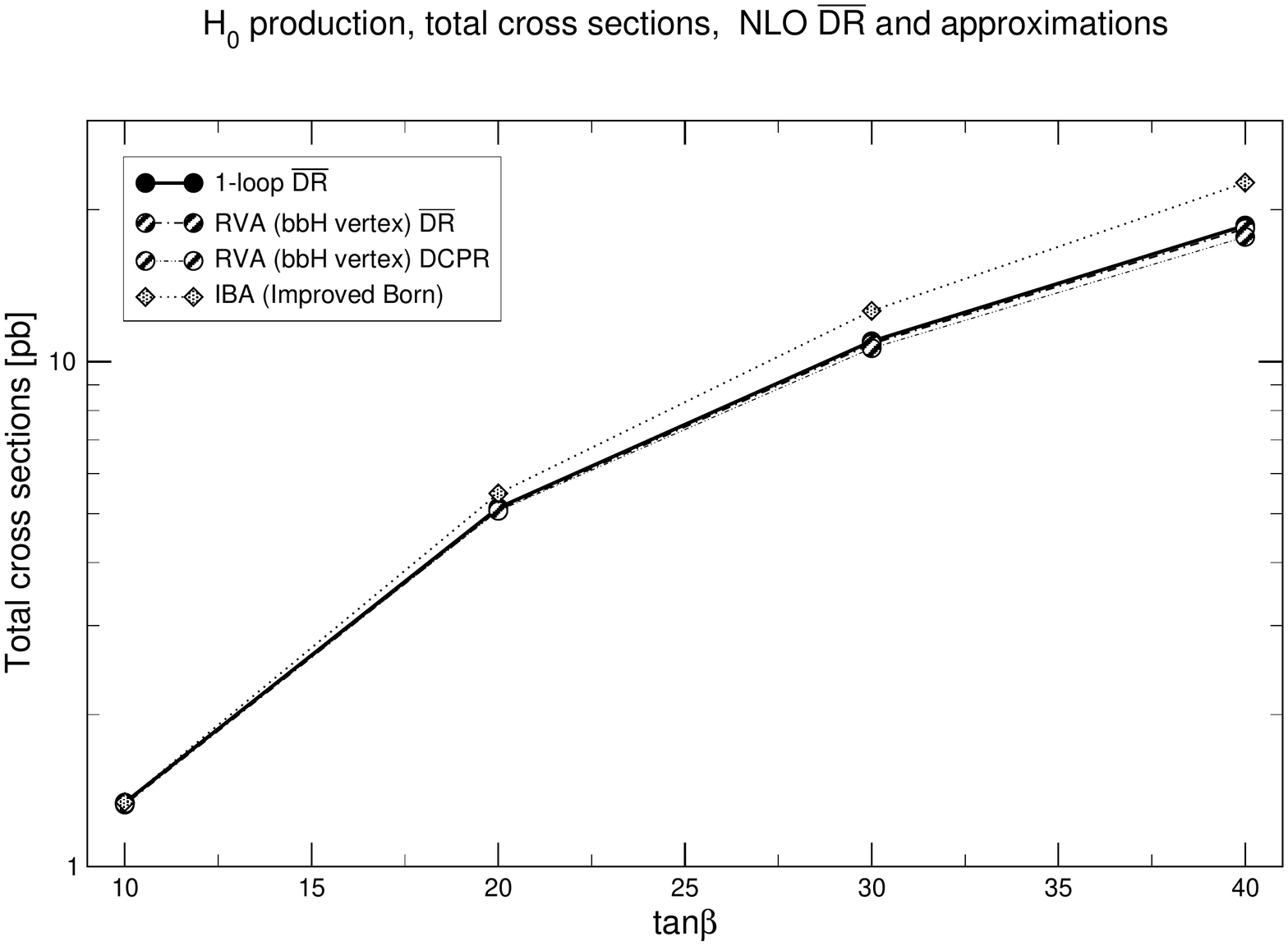}
\caption{Comparison of the total NLO cross sections: NLO \drbar, RVA (\drbar
  and DCPR) and Improved Born Approximantion (IBA), $H^0$
  production; $M_{A^0}=250$ GeV, $p_{b,T}>20$ GeV, $|y_b|<2$.
\label{fig:2IBA}}
\end{figure}
\begin{figure}
\centering
\includegraphics[width=1.0\textwidth]{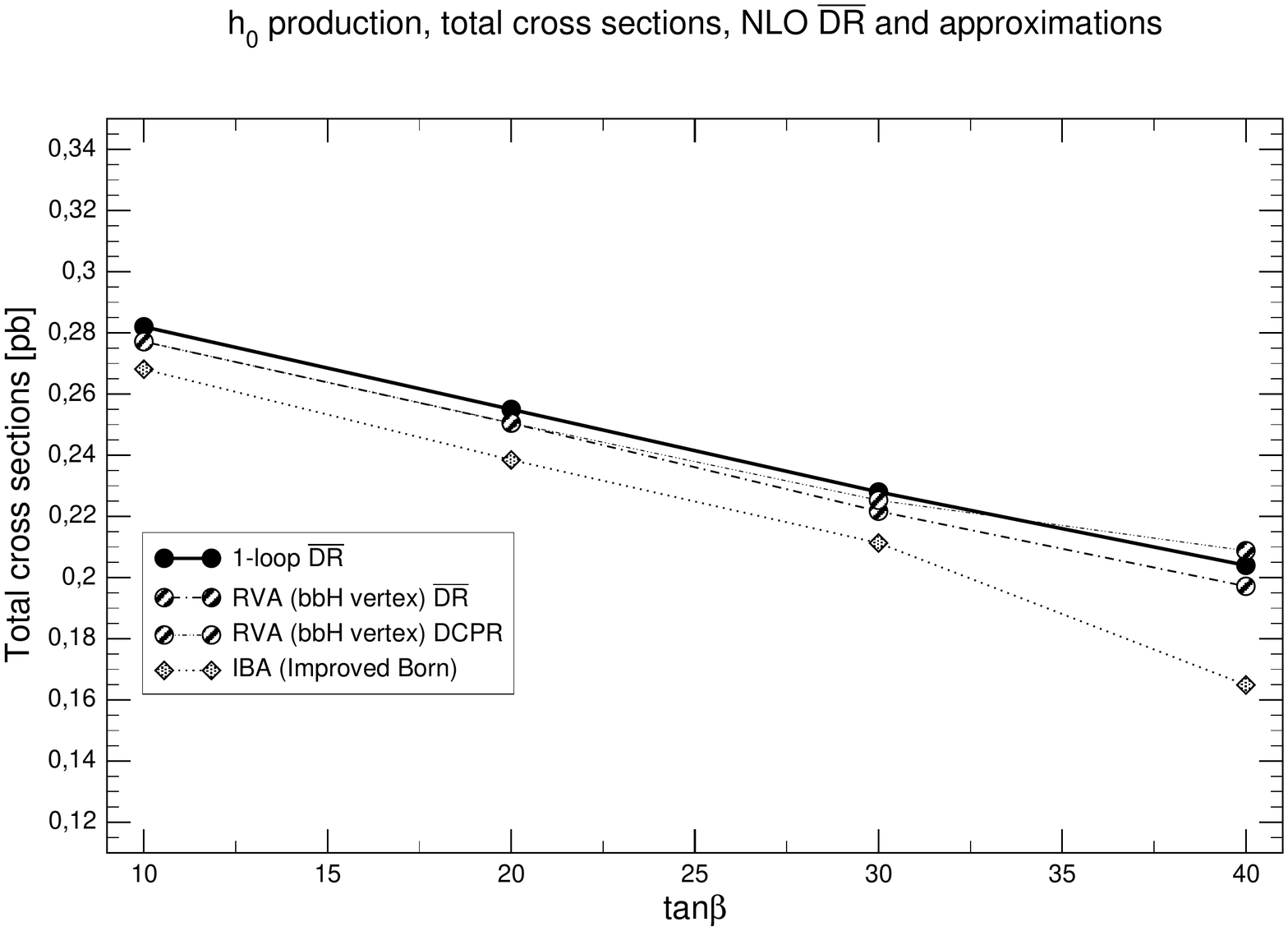}
\caption{Comparison of the total NLO cross sections: NLO \drbar, RVA (\drbar
  and DCPR) and Improved Born Approximantion (IBA), $h^0$
  production; $M_{A^0}=250$ GeV, $p_{b,T}>20$ GeV, $|y_b|<2$.
\label{fig:3IBA}}
\end{figure}

As one can see from the plots,
(Figures~\ref{fig:1IBA},\ref{fig:2IBA},\ref{fig:3IBA}) this version of IBA is sufficiently close to
the complete calculation only for relatively small $\tan \beta$ values, roughly
$\tan\beta < 20$. In this range, the approximation gives larger (compared to the
complete calculation) rates for $H^0,\, A^0$ and smaller rates for $h^0$. 
The differences remain below the ten percent size,
which would be tolerable at least in a first phase of LHC
measurements. Increasing the $\tan\beta$ value, the IBA description becomes 
worse. For $\tan\beta = 40$, it differs in all the three cases by, roughly, a
relative 25 percent, 
which seems a rather poor prediction for the measurable total rates.\\

For what concerns the $\tan\beta$ dependence of the plots, one can conclude
that it provides those features that would be expected at the chosen value
of $M_{A^0}$, which is sufficiently larger than $M_Z$ to approach the correct
“decoupling” limits. In this large $M_{A^0}$ regime, that is discussed widely 
in the literature (see e.g.~\cite{Djouadi:2005gj}), the 
 $H^0$ and $A^0$ couplings become almost exactly
proportional to $\tan\beta$, while the $h^0$ coupling becomes very weakly
$\tan\beta$ dependent. These features are well reproduced by the plots,
that show a roughly quadratic $\tan\beta$ dependence of the $H^0,\;A^0$ rates and a
much weaker $\tan\beta$ dependence for $h^0$. But for large $\tan\beta$ values,
there seems to be an extra $\tan\beta$ dependence of the complete calculation
that is not contained in the IBA description.\\


Having this apparent discrepancy in our mind, as a second attempt, we have tried to use what we would call a ``Reduced
Vertex Approximation'' (RVA): 
we approximate the complete NLO 
keeping only the (all) one loop corrections to the ``final'' Yukawa $bb\mathcal{H}^0$ vertex
and the subset of counterterms needed to get a UV-finite result; the photon mass is
regulated (arbitrarily) as $M_{\gamma} = M_Z$ (and thus we do not include soft and
hard radiation). We kept the one loop Higgs masses in the kinematics as well
as the $Z$-factors in the definition of the cross section; all the other
diagrams (Boxes, Initial and Up Triangles, Self Energies) are neglected. 
As a check we computed the cross section in this approximation 
in both schemes (the subset of diagrams, with the right choice of
counterterms, should be scheme independent). As one can see from
the updated figures our RVA turns out to provide very efficient
description of the total NLO cross sections; the difference between the NLO 
and the RVA is of order of $1\%$, $3.4 \%$ in the worst case. This is
numerically summarized in Tables~\ref{tab:IBA1},\ref{tab:IBA2},\ref{tab:IBA3}
and Figures~\ref{fig:1IBA},\ref{fig:2IBA},\ref{fig:3IBA}.\\

From the inspection of those Tables and Figures we
would conclude that the extra vertices that the RVA contains seem to
provide the extra $\tan\beta$ dependence not predicted by the IBA in a
reasonably satisfactory way, i.e. at the level of few percent in the full
$\tan\beta$ range. This RVA cannot be transformed into simple analytical
expressions. It tells us that the relative effect of a large set of Feynman
diagrams, those that were not included in the approximation, is small, at
the level of a few percent, which might be considered negligible for the
first phase of LHC measurements.
\begin{table}
\small{
\begin{tabular}{c|c|c|c|c|c}
\hline
$\tan \beta$  \quad  &  $\;\sigma^{\mbox{\tiny{\drbar,NLO}}}\;$ &
\quad RVA$_{bb\mathcal{H}}$ \quad  &  \quad $\sigma/$RVA$_{bb\mathcal{H}}$ \quad &  \quad IBA \quad &
 \quad $\sigma/$IBA \quad \\
\hline
10 & 1.338  & 1.32623& 1.00888& 1.34087 & 0.997861\\
20 & 5.133  & 5.08324& 1.00979& 5.48397 & 0.936\\
30 & 10.975 & 10.8433& 1.01215& 12.6044 & 0.87073\\
40 & 18.613 & 18.3461& 1.01455& 22.6229 & 0.822749\\
\hline
\end{tabular}
\caption{$H^0$ production: comparison between the complete NLO prediction and
  the two approximations: total cross sections and ratios 
  $\;\sigma^{\mbox{\tiny{\drbar,NLO}}}/\sigma_{\textrm{\tiny{APP.}}}$.}\label{tab:IBA1}}
\end{table}

\begin{table}
\small{
\begin{tabular}{c|c|c|c|c|c}
\hline
$\tan \beta$  \quad  &  $\;\sigma^{\mbox{\tiny{\drbar,NLO}}}\;$ &
\quad RVA$_{bb\mathcal{H}}$ \quad  &  \quad $\sigma/$RVA$_{bb\mathcal{H}}$ \quad &  \quad IBA \quad &
 \quad $\sigma/$IBA \quad \\
\hline
10 & 0.282& 0.277157& 1.01747& 0.268161& 1.05161\\
20 & 0.255& 0.250495& 1.01799& 0.238459& 1.06936\\
30 & 0.228& 0.221673& 1.02854& 0.211275& 1.07916\\
40 & 0.204& 0.197159& 1.0347&  0.164874& 1.23731\\
\hline
\end{tabular}
\caption{$h^0$ production: comparison between the complete NLO prediction and
  the two approximations: total cross sections and ratios
  $\;\sigma^{\mbox{\tiny{\drbar,NLO}}}/\sigma_{\textrm{\tiny{APP.}}}$.}\label{tab:IBA2}}
\end{table}

\begin{table}
\small{
\begin{tabular}{c|c|c|c|c|c}
\hline
$\tan \beta$  \quad  &  $\;\sigma^{\mbox{\tiny{\drbar,NLO}}}\;$ &
\quad RVA$_{bb\mathcal{H}}$ \quad  &  \quad $\sigma/$RVA$_{bb\mathcal{H}}$ \quad &  \quad IBA \quad &
 \quad $\sigma/$IBA \quad \\
\hline
10 & 1.367&  1.35328& 1.01014& 1.36737 & 0.999729\\
20 & 5.04&   4.98026& 1.01199& 5.4543  & 0.924042 \\
30 & 10.601& 10.4581& 1.01366& 12.2948 & 0.862232 \\
40 & 17.118& 16.8292& 1.01716& 21.7326 & 0.787663\\
\hline
\end{tabular}
\caption{$A^0$ production: comparison between the complete NLO prediction and
  the two approximations: total cross sections and ratios
  $\;\sigma^{\mbox{\tiny{\drbar,NLO}}}/\sigma_{\textrm{\tiny{APP.}}}$.}\label{tab:IBA3}}
\end{table}

\section{Conclusions}
\label{sec:concl}

We have performed in this paper a complete MSSM calculation of the electroweak NLO
effect in the processes of semi-inclusive bottom-Higgs production. 
Our analysis has been performed for two choices of the $M_{A^0}$ input 
parameter and for variable values of the $\tan\beta$ parameter defined in the
\drbar renormalization scheme. Although a more extended  analysis of the parameter space would
be interesting, we have found certain results that appear to us to be general 
and worth publishing. 
The first conclusion is that two different renormalization schemes appear 
to be practically identical at SUSY NLO as one would a priori expect. 
Working in the \drbar scheme, that seemed to us to be somehow preferable, 
we have found that the pure electroweak one-loop effect in the three
considered production processes is of a size that might be relevant
and therefore that this
contribution cannot be ignored for a proper experimental analysis of the
reactions.\\ 

There could exist simpler calculations involving a smaller (but still
large) number of diagrams, that would provide a valid numerical result. We
have seen that one possible Improved Born Approximation does not reproduce the correct
result in a satisfactory way. We have also seen that another ``Reduced
Vertex Approximation''
(which considers only the 1-loop correction to the Yukawa
$bb\mathcal{H}^0$ vertex) 
appears to better approximate the full NLO.\\ 

However, if a theoretical prediction of the
total cross section is requested at the percent level, which
might be the hopefully desirable final LHC goal, our conclusion is that the
complete one-loop calculation of the electroweak part that we have performed in this paper should
be considered, together with the available, large, QCD corrections, 
as the correct proposal to offer to the experimental
community.\\

There remains a couple of relevant points to be
still investigated. The first  is that of combining 
this analysis with an analogous one to be performed for the process of 
associated top-charged Higgs production, for which our group has already 
provided a complete one-loop electroweak analysis~\cite{Beccaria:2009my}. 
The second one is that of trying to relate the \drbar $\tan \beta$ parameter, 
which is not a measurable quantity, to a measurable $\tan \beta$ (which
could be defined for instance 
 by  $A^0 \to \tau^+ \tau^-$ decay as suggested in~\cite{schemes}). 
This would allow to draw plots where also the horizontal axis represents a 
measurable quantity. 
These points are, in our opinion, quite relevant but beyond the purposes of
our analysis; work is in progress on these issues.

\section*{Acknowledgements}

We want to thank A. Djouadi for several fruitful discussions.
E.~M. would like to thank Heidi~Rzehak and Jianhui~Zhang
for valuable comments and suggestions.  E.~M. is supported by the European
Research Council under Advanced Investigator Grant ERC-AdG-228301.

\newpage
\appendix

\section{Renormalization constants in the Higgs sector}
\label{App:AA}
The renormalization constants of the wavefunction of the Higgs bosons $A^0, h^0, H^0$
and of the Goldstone boson $G^0$ are given by
\begin{eqnarray}
&~& \delta \bar Z_{H^0H^0}=\cos^2\alpha\delta Z_{H1}+
\sin^2\alpha\delta Z_{H2} , ~~~
 \delta \bar Z_{H^0h^0}=\sin\alpha\cos\alpha
(\delta Z_{H2}-\delta Z_{H1}), \nonumber \\
&~& \delta \bar Z_{h^0h^0}=\sin^2\alpha\delta Z_{H1}+
\cos^2\alpha\delta Z_{H2}, ~~~~ 
\delta \bar Z_{A^0A^0}=\sin^2\beta\delta Z_{H1}+
\cos^2\beta\delta Z_{H2}, \nonumber \\
&~&
\delta \bar Z_{G^0A^0}=\cos\beta\sin\beta
(\delta Z_{H2}-\delta Z_{H1}), ~~~ 
\delta \bar Z_{h^0A^0}=\delta \bar Z_{H^0A^0}=
\delta \bar Z_{h^0G^0}=\delta \bar Z_{H^0G^0}=0. \nonumber \\
\end{eqnarray}
The renormalization constants  for the $c^\eta(bb h^0)$ and for  the 
 $c^\eta(bb H^0)$ couplings  is obtained differentiating the tree-level
 expressions in Eq.~(\ref{eq:couplings}),
\begin{eqnarray}
\delta c^\eta(bbh^0) &=&  \left ( 
{\delta g\over g}+{\delta m_b\over m_b}
-{\delta M^2_W\over 2  M^2_W}
-{\delta\cos\beta\over\cos\beta} \right )  c^\eta(bbh^0), \nonumber \\
\delta c^\eta(bbH^0) &=&  \left ( 
{\delta g\over g}+{\delta m_b\over m_b}
-{ \delta M^2_W\over 2  M^2_W}
-{\delta\cos\beta\over\cos\beta} \right )  c^\eta(bbH^0). 
 \end{eqnarray}
 $\delta \cos \beta$, $\delta M^2_W$, and $\delta g$, reads as  follows
\begin{eqnarray}
\delta\cos\beta &=& -\sin^2\beta{\delta\tan\beta\over\tan\beta}, \nonumber \\
\delta M^2_W &=& \mbox{Re}  \Sigma_{W}(M^2_W), \nonumber \\
{\delta g\over g}&=& {\Sigma_{\gamma Z}(0)\over s_Wc_W M^2_Z}
-\frac{1}{2} \Bigg [ 
+2{c_W\over s_W M^2_Z}\Sigma_{\gamma Z}(0)
+{c^2_W\over s^2_W} \left ( {\delta M^2_Z\over M^2_Z} -
{\delta M^2_W\over M^2_W} \right ) - \Sigma'_{\gamma\gamma}(0) \Bigg],
\end{eqnarray}
with $\delta M^2_Z =\mbox{Re}  \Sigma_{Z}(M^2_Z)$.  The $c^\eta(bbA^0)$
couplings depends only on the angle $\beta$. When computing the 
the renormalization constant $\delta c^\eta(bbA^0)$, one has 
to distinguish between the $\beta$-dependent factors originated by 
the $H_1$, $H_2$ mixing and the $\beta$- dependent factors  
from the $H_1$, $H_2$ couplings. Only the latter have to be renormalized. In
 particular the factor $\sin \beta$
[$1 / \cos \beta$] entering the $c^\eta(bbA^0)$  coupling is originated from the 
$H_1$, $H_2$ mixing [couplings], and thus $\delta c^\eta (bb A^0)$ reads
\bq
\delta c^\eta(bbA^0) =  \left ( 
{\delta g\over g}+{\delta m_b\over m_b}
-{\delta M^2_W\over 2  M^2_W}
-{\delta\cos\beta\over\cos\beta} \right )  c^\eta(bbA^0).
\eq  

\section{Contributions of the counterterms}
\label{app:A}
In this appendix we list explicitely the contributions of
the counterterms writen in terms of the renormalization constants introduced in 
Section~\ref{sub:C} and in Appendix~\ref{App:AA}.  The 
vertices counterterms 
can be written as follows 
\bq
\bar u'_b(\lambda'_b) \left ( \sum_{k=1}^4 \sum_{\eta =L,R} 
J^{k\eta} V^{k\eta}_{bg\to b \mathh^0}
 \right ) u_b(\lambda_b),
\eq
where $J^{k\eta}$ are defined in Eq.~(\ref{eq:JFF}) while the non-zero $V^{k\eta}_{bg \to b \mathh^0} $ 
reads
\begin{eqnarray}
V^{1\eta}_{bg \to b \mathh^0}  &=& 
{g_s\over s-m^2_b}
\left \{
\left ( {3\over2}\delta Z^b_{\eta}
+{1\over2} \delta Z^b_{\bar \eta} \right )c^{\eta}(bb\mathh^0)
+\delta c^{\eta}(bb\mathh^0)
+{1\over2}
\sum_{\bar \mathh^0 }\delta \bar Z^*_{ \bar \mathh^0   \mathh^0}c^{\eta}(bb \bar \mathh^0)   \right \}
\nonumber\\ 
&-& {g_s\over u-m^2_b }
\left \{ \left ( {3\over2}\delta Z^b_{\bar \eta}+
{1\over2}\delta Z^b_\eta \right  )c^{\eta}(bb  \mathh^0)
+\delta c^{\eta}(bb  \mathh^0)+{1\over2}
\sum_{\bar \mathh^0 }\delta \bar Z^*_{\bar \mathh^0 \mathh^0 }c^{\eta}(bb \bar \mathh^0) \right \},\nonumber\\
V^{2\eta}_{bg \to b \mathh^0}  &=&
{-2 g_s\over u-m^2_b }
\left \{ \left ( {3\over2}\delta Z^b_{\bar \eta}+
{1\over2}\delta Z^b_\eta \right  )c^{\eta}(bb  \mathh^0)
+\delta c^{\eta}(bb  \mathh^0)+{1\over2}
\sum_{\bar \mathh^0 }\delta \bar Z^*_{\bar \mathh^0 \mathh^0 }c^{\eta}(bb \bar \mathh^0) \right \} , \nonumber \\
V^{3\eta}_{bg \to b \mathh^0}  &=&
{m_b g_s\over s-m^2_b}
\left (\delta Z^b_{\bar \eta}-\delta Z^b_{\eta} \right )c^{\bar \eta}(bb  \mathh^0)
+ {m_b g_s\over u-m^2_b}
\left (\delta Z^b_{\bar \eta}-\delta Z^b_{\eta} \right )c^{ \eta}(bb  \mathh^0),
\end{eqnarray}
where $(\eta,  \bar \eta) \in \{ (L,R); \, (R,L) \}$ and $ \mathh^0,  \bar \mathh^0 = h^0 , H^0,  A^0, G^0$.
The bottom self energy counterterm reads as follows
\bq
\bar u'_b(\lambda'_b) \left ( \sum_{k=1}^4 \sum_{\eta =L,R} 
J^{k\eta} S^{k\eta}_{bg\to b \mathh^0}
 \right ) u_b(\lambda_b).
\eq
The non-zero $ S^{k\eta}_{bg\to b \mathh^0}$ are
\begin{eqnarray}
S^{1\eta}_{bg \to b \mathh^0}  &=& 
g_s
{c^{\eta}(bb\mathh^0)\over(s-m^2_b)^2}
\left \{ s \delta Z^b_{\eta}
- m^2_b\left (  \delta Z^b_{\eta}
- 2{\delta m_b\over m_b} \right ) \right  \}\nonumber\\
&+& g_s{c^{\eta}(bb\mathh^0) \over
(u-m^2_b)^2} \left \{ u \delta Z^b_{\bar \eta}
-m^2_b \left (
\delta Z^b_{\bar \eta} - 2 {\delta m_b\over m_b} \right ) \right \} \nonumber\\
S^{2 \eta}_{bg \to b \mathh^0}  &=& 2  g_s{c^{\eta}(bb\mathh^0) \over
(u-m^2_b)^2} \left \{ u \delta Z^b_{\bar \eta}
-m^2_b \left (
\delta Z^b_{\bar \eta} - 2 {\delta m_b\over m_b} \right ) \right \} \nonumber\\
S^{3 \eta}_{bg \to b \mathh^0}  &=&
g_s m_b
{c^{\bar \eta}(bb\mathh^0)\over(s-m^2_b)}
\left \{ {1\over2}
\left (\delta Z^b_{\eta}-\delta Z^b_{\bar \eta} \right )-~{\delta m_b\over m_b} \right \} \nonumber \\
&+& g_s m_b {c^{\eta}(bb\mathh^0)\over
(u-m^2_b)}\left \{ 
{1\over2} \left (\delta Z^b_{\eta}
-\delta Z^b_{\bar \eta} \right )-{\delta m_b\over m_b} \right \} 
\end{eqnarray}

\newpage



\newpage

\vspace{-1cm}


\clearpage




\begin{figure}
\centering
\includegraphics[width=1.0\textwidth]{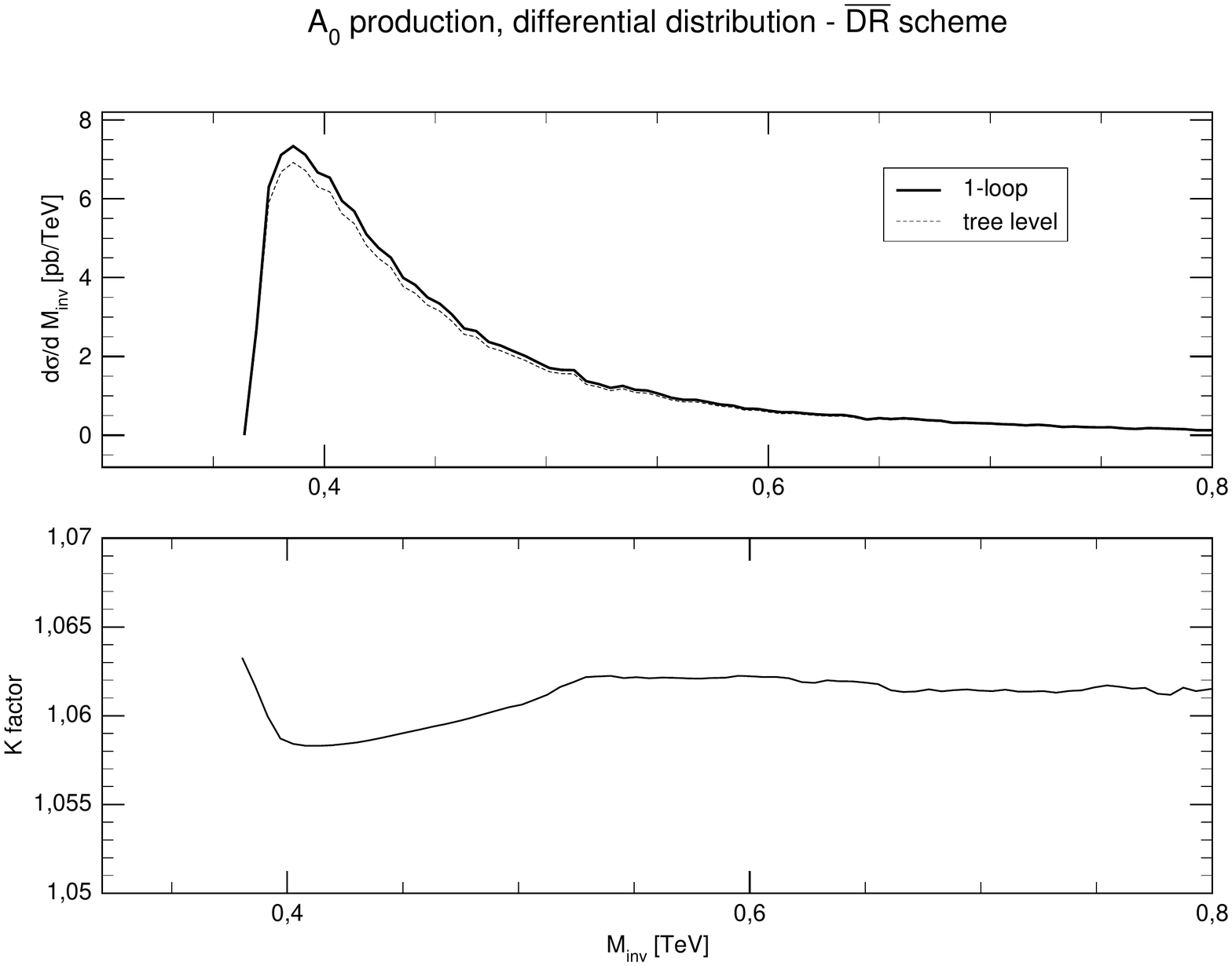}
\caption{Invariant mass distribution, $A^0$ production, \drbar scheme.  
$M_{A^0}=350$ GeV, $\tan \beta=15$ $p_{b,T}>20$ GeV, $|y_b|<2$.
\label{A0-MA350-tb15-K-QED}}
\end{figure}
\begin{figure}
\centering
\includegraphics[width=1.0\textwidth]{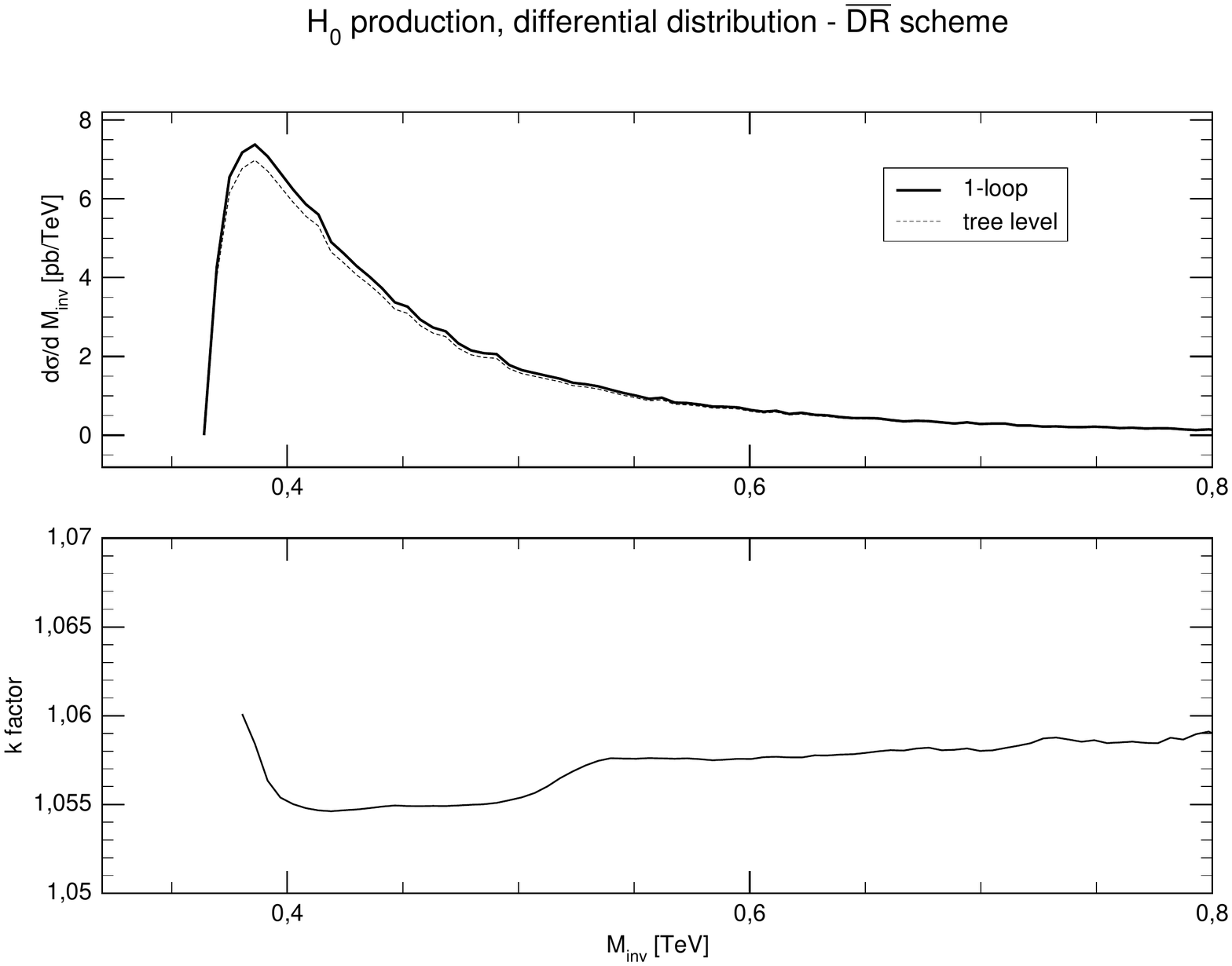}
\caption{Invariant mass distribution, $H^0$ production, \drbar scheme.  
$M_{A^0}=350$ GeV, $\tan \beta=15$ $p_{b,T}>20$ GeV, $|y_b|<2$.
\label{H0-MA350-tb15-K-QED}}
\end{figure}

\end{document}